\catcode`\@=11
\newif\if@fewtab\@fewtabtrue

{\count255=\time\divide\count255 by 60
\xdef\hourmin{\number\count255}
\multiply\count255 by-60\advance\count255 by\time
\xdef\hourmin{\hourmin:\ifnum\count255<10 0\fi\the\count255}}
\def\ps@draft{\let\@mkboth\@gobbletwo
    \def\@oddhead{}
    \def\@oddfoot
       {\hbox to 7 cm{$\scriptstyle Draft\ version:\ \draftdate$
       \hfil}\hskip -7cm\hfil\rm\thepage \hfil}
    \def\@evenhead{}\let\@evenfoot\@oddfoot}

\def\ceqno{\global\@fewtabfalse
    \ifcase\@eqcnt \def\@tempa{& & &}\or \def\@tempa{& &}
      \or \def\@tempa{&}
      \or\def\@tempa{}\fi\@tempa
{\rm(\theequation)}}

\def\aeqno#1{\global\@fewtabfalse
    \ifcase\@eqcnt \def\@tempa{& & &}\or \def\@tempa{& &}
      \or \def\@tempa{&}
      \or\def\@tempa{}\fi\@tempa
{\rm(\theequation,#1)}}

\def\label#1{\ifnum\draftcontrol=1
 \global\def\draftnote{$\scriptstyle #1$}\fi
 \@bsphack\if@filesw {\let\thepage\relax
   \def\protect{\noexpand\noexpand\noexpand}%
\xdef\@gtempa{\write\@auxout{\string
      \newlabel{#1}{{\@currentlabel}{\thepage}}}}}\@gtempa
   \if@nobreak \ifvmode\nobreak\fi\fi\fi
  \@esphack}

\def\alabel#1#2{\label{#1}\global\@fewtabfalse
    \ifcase\@eqcnt \def\@tempa{& & &}\or \def\@tempa{& &}
      \or \def\@tempa{&}
      \or\def\@tempa{}\fi\@tempa
{\hbox to 3cm{\phantom{\rm(\theequation,#2)}
\draftnote \hfil}\hskip -3cm {\rm(\theequation,#2)}}}

\def\clabel#1{\label{#1}\global\@fewtabfalse
    \ifcase\@eqcnt \def\@tempa{& & &}\or \def\@tempa{& &}
      \or \def\@tempa{&}
      \or\def\@tempa{}\fi\@tempa
{\hbox to 3cm{\phantom{\rm(\theequation)}
\draftnote \hfil}\hskip -3cm{\rm(\theequation)}}}

\def\eqnarray{\def\draftnote{{}}\global\@fewtabtrue
\stepcounter{equation}\let\@currentlabel=\theequation
\global\@eqnswtrue
\global\@eqcnt\z@\tabskip\@centering\let\\=\@eqncr
$$\halign to \displaywidth\bgroup\@eqnsel\hskip\@centering\@eqcnt\z@
  $\displaystyle\tabskip\z@{##}$&\global\@eqcnt\@ne
  \hskip 1\arraycolsep \hfil${##}$\hfil
  &\global\@eqcnt\tw@ \hskip 1\arraycolsep
$\displaystyle\tabskip\z@{##}$
\hfil  \tabskip\@centering&\global\@eqcnt\thr@@\llap{##}\tabskip\z@
\cr}

\def\endeqnarray{\@@eqncr\egroup
      \global\advance\c@equation\m@ne$$\global\@ignoretrue}

\def\@eqnnum{\hbox to 3cm{\phantom{\rm(\theequation)} \draftnote
                         \hfil}\hskip -3cm {\rm(\theequation)}}

\def\@@eqncr{\let\@tempa\relax
    \ifcase\@eqcnt \def\@tempa{& & &}\or \def\@tempa{& &}
      \or \def\@tempa{&}
      \or\def\@tempa{}
\fi\@tempa
\if@eqnsw
\if@fewtab\@eqnnum\fi
\stepcounter{equation}\fi\global
\@eqnswtrue\global\@eqcnt\z@\global\@fewtabtrue\cr}

\def\draftcite#1{\ifnum\draftcontrol=1#1\else{}\fi}

\def\@lbibitem[#1]#2{\item{}\hskip -3cm \hbox to 2cm
{\hfil$\scriptstyle\draftcite{#2}$}\hskip
1cm[\@biblabel{#1}]\if@filesw
     {\def\protect##1{\string ##1\space}\immediate
      \write\@auxout{\string\bibcite{#2}{#1}}}\fi\ignorespaces}

\def\@bibitem#1{\item\hskip -3cm \hbox to 2cm
{\hfil $\scriptstyle\draftcite{#1}$}\hskip 1cm
\if@filesw \immediate\write\@auxout
       {\string\bibcite{#1}{\the\value{\@listctr}}}\fi\ignorespaces}

\def\nsection#1{\section{#1}\setcounter{equation}{0}}

\def\nappendix#1{\vskip 1cm\no{\bf Appendix #1}\def\thesection{#1}
\setcounter{equation}{0}}

\font\tendl=msbm10  scaled \magstep1
\font\sevendl=msbm7 scaled \magstep1
\font\fivedl=msbm5 scaled \magstep1
\font\tengl=eufm10  scaled \magstep1
\font\sevengl=eufm7 scaled \magstep1
\font\fivegl=eufm5 scaled \magstep1

\newfam\dlfam \def\dl{\fam\dlfam\tendl} 
\textfont\dlfam=\tendl \scriptfont\dlfam=\sevendl
\scriptscriptfont\dlfam=\fivedl
\newfam\glfam  
\textfont\glfam=\tengl \scriptfont\glfam=\sevengl
\scriptscriptfont\glfam=\fivegl

\def\draftdate{\number\month/\number\day/\number\year\ \ \ \hourmin }

\global\def\draftcontrol{0}
\catcode`\@=12
\def\tilde{\widetilde}

\documentstyle[11pt]{article}

\renewcommand{\theequation}{\thesection.\arabic{equation}}
\def\theequation{{\thesection.\arabic{equation}}}

\newcommand{\be}{\begin{eqnarray}}
\newcommand{\en}{\end{eqnarray}\vs 0.5 cm}

\newcommand{\no}{\noindent}
\newcommand{\vs}{\vskip}
\newcommand{\hs}{\hspace}

\newcommand{\NP}{{{\dl P}}}
\newcommand{\NC}{{{\dl C}}}
\newcommand{\NZ}{{{\dl Z}}}
\newcommand{\Ng}{{\bf g}}
\newcommand{\qq}{\begin{eqnarray}}
\newcommand{\de}{\bar\partial}
\newcommand{\da}{\partial}
\newcommand{\ee}{{\rm e}}

\newcommand{\qqq}{\end{eqnarray}}

\newcommand{\CA}{{\cal A}}

\newcommand{\CG}{{\cal G}}
\newcommand{\CH}{{\cal H}}

\newcommand{\CK}{{\cal K}}

\newcommand{\CN}{{\cal N}}
\newcommand{\CO}{{\cal O}}

\newcommand{\CU}{{\cal U}}

\newcommand{\CZ}{{\cal Z}}
\newcommand{\s}{\hspace{0.05cm}}
\newcommand{\m}{\hspace{0.025cm}}

\newcommand{\hf}{{_1\over^2}}

\begin{document}
\title{\bf{Self-duality of the $SL_2$ Hitchin integrable\\ 
system at genus 2}}
\author{\ \\Krzysztof Gaw\c{e}dzki \\ I.H.E.S., C.N.R.S.,
F-91440  Bures-sur-Yvette, France\\ \\Pascal Tran-Ngoc-Bich \\
Universit\'{e} de Paris Sud, F-91405 Orsay, France}
\date{ }
\maketitle

\vskip 0.3cm
\vskip 1 cm

\begin{abstract}
\vskip 0.3cm

\noindent We revisit the Hitchin integrable system
\cite{Hitch}\cite{VGP} whose phase space is the bundle cotangent 
to the moduli space $\cal N$ of holomorphic $SL_2$-bundles 
over a smooth complex curve of 
genus 2. As shown in \cite{NarRa}, $\m\cal{N}$ 
may be identified with the 3-dimensional projective space of theta 
functions of the 2$^{\rm nd}$ order, i.e. $\CN\cong \NP^{3}$. 
We prove that the Hitchin system on $T^*\CN\cong T^*\NP^{3}$ 
possesses a remarkable symmetry: it is invariant under 
the interchange of positions and momenta. This property allows 
to complete the work of van Geemen-Previato \cite{VGP} which,
basing on the classical results on geometry of the Kummer quartic 
surfaces, specified the explicit form of the Hamiltonians
of the Hitchin system. The resulting integrable system resembles 
the classic Neumann systems which are also self-dual. 
Its quantization produces a commuting family of differential operators 
of the 2$^{\rm nd}$ order acting on homogeneous polynomials in four 
complex variables. As recently shown by van Geemen-de Jong 
\cite{VGDJ}, these operators realize 
the Knizhnik-Zamolodchikov-Bernard-Hitchin
connection for group $SU(2)$ and genus 2 curves.
\end{abstract}
\vs 2cm

\nsection{Introduction}
\setcounter{equation}{0}
\vskip 0.5cm

In \cite{Hitch}, Nigel Hitchin has discovered an interesting 
family of classical integrable models related to modular
geometry of holomorphic vector bundles or to 2-dimensional
gauge fields. The input data for Hitchin's construction
are a complex Lie group $G$ and a complex  curve $\Sigma$ 
of genus $\gamma$. The configuration space of the integrable 
system is the moduli space $\cal N$ of (semi)stable holomorphic 
$G$-bundles over $\Sigma$. This is a finite-dimensional 
complex variety and Hitchin's construction is done in the holomorphic
category. It exhibits a complete family of Poisson-commuting 
Hamiltonians on the (complex) phase space $T^*\cal{N}$.  
The Hitchin Hamiltonians have open subsets of abelian varieties 
as generic level sets on which they induce additive
flows \cite{Hitch}. More recently, Hitchin's construction 
was extended to the case of singular or punctured curves 
\cite{Mark}\cite{Nekr}\cite{EnR} providing a unified construction 
of a vast family of classical integrable systems. For $\Sigma=\NC P^1$ 
with punctures, one obtains this way the so called Gaudin chains
and for $G=SL_N$ and $\Sigma$ of genus 1 with one puncture,
the elliptic Calogero-Sutherland models which
found an unexpected application in the 
supersymmetric 4-dimensional gauge theories \cite{DonaW}.
\vskip 0.4cm

In Section 2 of the present paper we briefly
recall the basic idea of Hitchin's construction. The main  
aim of this contribution is to treat in detail the
case of $G=SL_2$ and $\Sigma$ of genus 2 (no punctures).
The genus 2 curves are hyperelliptic, i.e.\s\s given
by the equation
\qq
\zeta^2\s=\s\prod\limits_{s=1}^6(\lambda-\lambda_s)
\label{hellp}
\qqq
where $\lambda_s$ are 6 different complex numbers.
The semistable moduli space $\CN$ has a particularly simple form 
for genus 2, \cite{NarRa}: it is the
projectivized space of theta functions of the 2$^{\rm nd}$ 
order:
\qq
\CN\ =\ \NP H^0(L_{\Theta}^2)
\label{be}
\qqq
where $L_\Theta$ is the theta-bundle over the Jacobian $J^1$
of (the isomorphism classes of) degree 
$\gamma-1=1$ line bundles\footnote{we use
the multiplicative notation for the tensor product of line 
bundles} $l$ over $\Sigma$. \s${dim}_{_{\bf C}}(H^0(L_\Theta^2))=4$
so that $\CN\cong\NP^3$. This picture of $\CN$ is related 
to the realization of $SL_2$-bundles as extensions 
of degree 1 line bundles. We review some 
of the results in this direction in Section 3 using 
a less sophisticated language than that of the original
work \cite{NarRa}. The relation between the extensions 
and the theta functions is lifted to the level of the cotangent 
bundle $T^*\CN$ in Section 4. The language of extensions 
proves suitable for a direct description of the Hitchin Hamiltonians 
on $T^*\CN$. The main aim is, however, to present the Hitchin system 
as an explicit 3-dimensional family of integrable systems 
on $T^*\NP^3$, parametrized by the moduli of the curve.
This was first attempted, and almost achieved, in reference 
\cite{VGP}.
\vskip 0.4cm

Let us recall that the Hitchin Hamiltonians 
are components of the map
\qq
\CH\m:\ T^*\CN\ \longrightarrow\ H^0(K^2)
\label{Hm}
\qqq
with values in the (holomorphic) quadratic differentials
($K$ denotes the canonical bundle of $\Sigma$).
Due to relation (\ref{be}), the map $\CH$ may be viewed as 
a $H^0(K^2)$-valued function of pairs $(\theta,\phi)$ where 
$\theta\in H^0(L_{\Theta}^2)$ and $\phi$ from the dual space
$H^0(L_\Theta^2)^*$ are s.t. ${\langle\m}\theta,
\m\phi{\m\rangle}=0$. Fix a holomorphic trivialization 
of $L_\Theta$ around $l\in J^1$ and denote by $\phi_l$ 
the linear form that computes the value of the theta function 
at $l$. As was observed in \cite{VGP}, 
\qq
\CH(\theta,\m\phi_l)\ =\ -\m{_1\over^{16\pi^2}}
\s(d\theta(l))^2
\label{onK}
\qqq
(with appropriate normalizations). In the above   
formula, $\theta$ is viewed as a function
on $J^1$ and $d\theta(l)$ as an element of $H^0(K)$.
Since $\theta(l)=0$, the equation is consistent with 
changes of the trivialization of $L_\Theta$.
\vskip 0.4cm

The map $J^1\ni l\mapsto\phi_l$ induces
an embedding of the Kummer surface $J^1/\NZ_2$
with $l$ and $l^{-1}K$ identified into a quartic
$\CK^*$ in $\NP H^0(L_\Theta^2)^*$. The Kummer quartic
is a carrier of a rich but classical structure, a subject 
of an intensive study of the nineteenth century geometers, 
see \cite{Kummer} and also the last chapter of \cite{GH}. 
The reference \cite{VGP} used the relation (\ref{onK}) and 
a mixture of the classical results and of more modern
algebraic geometry to recover an explicit form of the components
of the Hitchin map $\CH$ up to a multiplication by a function on 
the configuration space. The authors of \cite{VGP} checked
that the simplest way to fix this ambiguity leads 
to Poisson-commuting functions but they fell short of showing 
that the latter coincide with the ones of the Hitchin 
construction.
\vskip 0.4cm

Among the aims of the present paper is to fill the 
gap left in \cite{VGP}. We observe that the proposal 
of \cite{VGP} has a remarkable {\bf self-duality} property: 
it is invariant under the interchange of the positions 
and momenta in $T^*\NP^3$. We show that the Hitchin 
construction leads to a system with the same symmetry. This limits
the ambiguity left by the analysis of \cite{VGP} to a multiplication
of the components of $\CH$ by constants. A direct check based on
Eq.\s\s(\ref{onK}) fixes the normalizations and results 
in a formula for the Hitchin map which uses the hyperelliptic 
description (\ref{hellp}) of the curve. Namely,
\qq 
\CH\ =\ -\m{_1\over^{128\m\pi^2}}\s\sum\limits_{1\leq s\not=t\leq 6}
{r_{st}\over(\lambda-\lambda_s)(\lambda-\lambda_t)}\,(d\lambda)^2
\label{GR}
\qqq
where $r_{st}$ are explicit polynomials in $(\theta,\phi)$
given, upon representation of $(\theta,\phi)$ by
pairs $(q,p)\in\NC^4\times\NC^4$, by Eqs.\s\s(\ref{rs})
below. The above expression for $\CH$ has a similar form as that  
for the Hitchin map on the Riemann sphere with 
6 insertion points $\lambda_s$, see e.g. Sect. 4 of \cite{FalKG}, 
except for the structure of the terms $r_{st}$. This is not 
an accident but is connected to the reduction of conformal field 
theory on genus 2 surfaces to an orbifold theory in genus 0
\cite{Knizh}\cite{Zamo}. We plan to return to this relation 
in a future publication.
\vskip 0.4cm

Let us discuss in more details how we establish the 
self-duality of the Hitchin Hamiltonians. The main tool 
here is an explicit expression for the values of the 
Hitchin map off the Kummer quartic $\CK^*$ which
we obtain in Section 5. Our formula 
for $\CH(\theta,\phi)$ requires a choice 
of a pair of perpendicular 2-dimensional
subspaces $(\Pi,\Pi^\perp)$ where $\theta\in\Pi\subset
H^0(L_\Theta^2)$ and $\phi\in\Pi^\perp\subset H^0(L_\Theta^2)^*$
(there is a complex line of such choices). 
The plane $\Pi^\perp$ corresponds to a line $\NP\Pi^\perp$ 
in $\NP H^0(L_\Theta^2)^*$ which intersects the Kummer quartic 
$\CK^*$ in four points $\NC^*\phi_{l_j},\ j=1,2,3,4,$ (counting
with multiplicity). Whereas the analysis of \cite{VGP}
was mainly concerned with the geometry of bitangents
to $\CK^*$ with two pairs of coincident $\phi_{l_j}$'s,
we concentrate on the generic situation with $\phi_{l_j}$'s 
different. Then any two of them, say $\NC^*\phi_{l_1}$ 
and $\NC^*\phi_{l_2}$, span $\Pi^\perp$. $\Pi$
is composed of the 2$^{\rm nd}$ order theta functions vanishing 
at $l_1$ and $l_2$. In particular, 
\qq
\phi\s=\s a_1\m\phi_{l_1}\m+\m a_2\m\phi_{l_2}\quad
{\rm and}\quad\theta(l_1)\s=\s0\s=\s\theta(l_2)\s.
\label{ss0}
\qqq
Let $x_1+x_2$ and $x_3+x_4$ be the divisors of $l_1l_2$ and
of $l_1l_2^{-1}K$, respectively, where $x_i$ are four 
points\footnote{the other two lines of intersection 
of $\NP\Pi^\perp$ with $\CK^*$ correspond 
to $l_3$ and $l_4$ with $l_1l_3=\CO(x_1+x_3)$,
$l_1l_3^{-1}K=\CO(x_2+x_4)$, $l_1l_4=\CO(x_1+x_4)$,
$l_1l_4^{-1}K=\CO(x_2+x_3)$} in $\Sigma$.
If $l_1^2\not=K$, which holds in a general situation,
then the quadratic differential $\CH(\theta,\m\phi)$
is determined by its values at $x_i$ which, as we show
in Section 5, are
\qq
\CH(\theta,\m\phi)(x_i)\ = \ 
-\m {_1\over^{16\pi^2}}\s\left(a_1\s d\theta(l_1)\pm a_2\s 
d\theta(l_2)\right)^2(x_i)\s.
\label{fr}
\qqq
Sign plus is taken for $x_1$ and $x_2$
and sign minus for $x_3$ and $x_4$. Note that for $\phi=\phi_l$ 
with $\theta(l)=0$ the above equation reproduces
the result (\ref{onK}).
\vskip 0.4cm

As we recall at the end of Section 3, there exists an almost natural 
linear isomorphism $\iota$ between $H^0(L_\Theta^2)^*$ 
and $H^0(L_\Theta^2)$. What follows is independent of the remaining 
ambiguity in the choice of $\iota$. The identity 
$\langle\m\theta,\phi\m\rangle=\langle\m
\iota(\phi),\m\iota^{-1}(\theta)\m\rangle$ implies that
if $(\theta,\phi)$ is a perpendicular pair then so
is $(\theta',\m\phi')$ where $\theta'=\iota(\phi)$ and
$\phi'=\iota^{-1}(\theta)$.  Thus $\iota$ 
interchanges the positions and momenta in $T^*\CN$. 
We may take $(\Pi',\m{\Pi'}^\perp) 
=(\iota(\Pi^\perp),\m\iota^{-1}(\Pi))$ as a pair of perpendicular 
subspaces containing $(\theta',\phi')$. The line 
$\NP{\Pi'}^\perp$ meets $\CK^*$ in four points 
$\NC^*\phi_{l'_j}$. Equivalently, 
$\NC^*\iota(\phi_{l'_j})$ are the points of intersection of $\NP\Pi$ 
with the Kummer quartic $\CK=\iota(\CK^*)\subset\NP H^0(L_\Theta^2)$. 
In general situation, ${\Pi'}^\perp$ is spanned by any pair 
of $\phi_{l'_j}\m$'s so that 
\qq
\phi'\s=\s a'_1\m\phi_{l'_1}\m+\m a'_2\m\phi_{l'_2}\quad{\rm and}
\quad\theta'(l'_1)\s=\s0\s=\s\theta'(l'_2)
\label{ss2}
\qqq
which is the dual version of relations (\ref{ss0}).
Equivalently,
\qq
\theta\s=\s a'_1\s\iota(\phi_{l'_1})\s+\s a'_2\s\iota(\phi_{l'_2})
\quad{\rm and}\quad{\langle\m}\iota(\phi_{l'_1}),
\s\phi{\m\rangle}\s=\s0\s=\s
{\langle\m}\iota(\phi_{l'_2}),\s\phi{\m\rangle}\s.
\label{ss1}
\qqq
Let $y_i$ be the points associated 
to $l'_j$ the same way as the points $x_i$ were associated 
to $l_j$.  \s$l'_j$ may be chosen so that $y_i$ and $x_i$ coincide 
modulo the natural involution of $\Sigma$ fixing the six 
Weierstrass points. Formula (\ref{fr}) implies then that
\qq
\CH(\theta',\m\phi')(y_i)\ = \ 
-\m {_1\over^{16\pi^2}}\s\left(a'_1\s d\theta'(l'_1)\pm a'_2\s 
d\theta'(l'_2)\right)^2(y_i)\s.
\label{frd}
\qqq
Points $y_i$ in Eq.\s\s(\ref{frd}) may be replaced by $x_i$ 
since the quadratic differentials are equal at point $x$ 
if and only if they are equal at the image of $x$ by the
involution of $\Sigma$. A direct calculation of 
the coefficients $a_1,\s a_2$ and $a'_1,\s a'_2$ appearing 
on the right hand sides of Eqs.\s\s(\ref{fr})
and (\ref{frd}) shows then that both expressions coincide,
establishing the self-duality of $\CH$. The verification 
of this equality is the subject of Section 6.
\vskip 0.4cm

In Section 7, we recall the main result of reference \cite{VGP}
and show how the self-duality may be used to complete
the analysis performed there and to obtain the explicit form 
(\ref{GR}) of the Hitchin map. We briefly discuss the relation
of that form to the classical Yang-Baxter equation.
\vskip 0.4cm

An appropriate quantization of Hitchin Hamiltonians leads to operators 
acting on holomorphic sections of powers of the determinant line bundle 
over $\CN$ and defining the Knizhnik-Zamolodchikov-Bernard-Hitchin
\cite{KZ}\cite{Ber}\cite{Ber2}\cite{Hitch2} connection.
In our case, the sections of the powers
of the determinant bundle are simply homogeneous
polynomials on $H^0(L_\Theta^2)$. It is easy to quantize
the Hamiltonians corresponding to the components of 
the Hitchin map (\ref{GR}) in such a way that one obtains
an explicit family of commuting 2$^{\rm nd}$ order differential
operators acting on such polynomials. The corresponding connection 
coincides with the explicit form of the (projective) KZBH connection
worked out recently\footnote{we thank B. van Geemen for attracting
our attention to ref. \cite{VGDJ} and for pointing out that this work
may be used to fix indirectly the precise form of the Hitchin map}
in \cite{VGDJ}. 
\vskip 0.4cm

The quantization of the genus 2 Hitchin system
is briefly discussed in Conclusions, where we also mention other possible
directions for further research. Four appendices which close the paper 
contain some of more technical material.
\vskip 0.4cm

We would like to end the presentation of our paper
by expressing some regrets. We apologize to Ernst Eduard Kummer 
and other nineteenth century giants for our insufficient 
knowledge of their classic work. The apologies
are also due to few contemporary algebraic geometers who could
be interested in the present work for an analytic character
of our arguments. To the specialist in integrability
we apologize for the yet incomplete analysis of the integrable 
system studied here and, finally, we apologize to ourselves 
for not having finished this work 2 years ago. 
\vskip 0.9cm

\nsection{Hitchin's construction}
\vskip 0.5cm

Let us assume, for simplicity, that the complex Lie group $G$ is simple,
connected and simply connected. We shall denote by $\Ng$
its Lie algebra. The complex curve $\Sigma$ will be 
assumed smooth, compact and connected. Topologically,
all $G$-bundles on $\Sigma$ are trivial and the complex 
structures in the trivial bundle may be described by giving
operators $\de+A$ where $A$ are
smooth $\Ng$-valued 0,1-forms on $\Sigma$ \cite{AtiyBott}.
Let $\cal{A}$ denote the space of such forms 
(i.e. of chiral gauge fields). The group $\cal G$ of 
local (chiral) gauge transformations composed of smooth maps 
$h$ from $\Sigma$ to $G$ acts on operators $\de+A$ 
by conjugation and on the gauge fields $A$ by
\qq
A\ \longmapsto\ {}^h\hspace{-0.14cm}A\s\equiv\s 
h\s A\s h^{-1}\s+\s h\s \de h^{-1}\s.
\nonumber
\qqq
Two holomorphic $G$-bundles are equivalent iff the
corresponding gauge fields are in the same orbit of $\CG$.
Hence the space of orbits $\CA/\CG$ coincides with
the (moduli) space of inequivalent holomorphic $G$-bundles.
It may be supplied with a structure of a variety provided one
gets rid of bad orbits. This may be achieved by limiting 
the considerations to (semi)stable 
bundles, i.e. such that the vector bundle associated 
with the adjoint representations of $G$ contains only 
holomorphic subbundles with negative (non-positive) 
first Chern number. For $\gamma>1$, the moduli space 
$\CN_s\equiv\CA_s/\CG$ 
of stable $G$-bundles is a smooth complex variety with 
a natural compactification to a variety $\CN_{ss}$, 
the (Seshadri-)moduli space of semistable bundles \cite{NarRa}.
\vskip 0.4cm   

The complex cotangent bundle $T^*\CN_s$ may be obtained
from the infinite-dimensional bundle $T^*\CA_s$ by 
the symplectic reduction. $T^*\CA_s$ may be realized as
the space of pairs $(A,\Phi)$ where $\Phi$
is a (possibly distributional) $\Ng$-valued 1,0-form on $\Sigma$,
$A\in\CA_s$ and the duality with the vectors 
$\delta A$ tangent to $\CA$ is given by
\qq
\int_{_\Sigma} tr\ \Phi\wedge\delta A
\nonumber
\qqq
with $tr$ standing for the Killing form.
The action of the local gauge group $\CG$ on $\CA_s$
lifts to a symplectic action on $T^*\CA_s$ by
\qq
\Phi\ \longmapsto\ {}^h\hs{-0.05cm}\Phi\s\equiv\s 
h\s\Phi\s h^{-1}\ .
\nonumber
\qqq
The moment map $\mu$ for the action of $\CG$ on $T^*\CN_s$
is
\qq
\mu(A,\Phi)\ =\ \de\Phi+A\wedge\Phi
+\Phi\wedge A\ \equiv\ \de_{_{A}}\Phi\ .
\nonumber
\qqq
Note that it takes values in $\Ng$-valued 2-forms on $\Sigma$.
These may be naturally viewed as elements of the space 
dual to the Lie algebra of $\CG$. The symplectic reduction
of $T^*\CA_s$ realizes $T^*\CN_s$ as the
space of $\CG$-orbits in the zero level of $\mu$:
\qq
T^*\CN_s\ \cong\ \mu^{-1}(\{0\})\s/\s\CG\ .
\nonumber
\qqq
\vskip 0.4cm

For a homogeneous $G$-invariant polynomial $P$ on $\Ng$ of
degree $d_P$, the gauge invariant expression \s$P(\Phi)$ 
defines a section of the bundle $K^{d_P}$ of $d_P$-differentials 
on $\Sigma$. If $\Phi$ is in the zero level of $\mu$
then \s$P(\Phi)$ is also holomorphic. Hence the map 
\s$\Phi\s\mapsto\s P(\Phi)\s$ induces a map 
\qq
\CH_P\s:\ T^*\CN_s\ \longrightarrow\ H^0(K^{d_P})
\nonumber
\qqq
into the finite dimensional vector space of holomorphic 
differentials of degree $d_P$ on $\Sigma$. The components
of such vector-valued Hamiltonians clearly Poisson-commute 
since upstairs (on $T^*\CA_s$) they depend only 
on the momentum variables $\Phi$. By a beautiful
argument, Hitchin showed \cite{Hitch} that taking 
all polynomials $P$ one obtains a complete system 
of Hamiltonians in involution and that the 
collection of maps $\CH_P$ defines in generic 
points a foliation of $T^*\CN_s$ into (open subsets of)
abelian varieties.
\vskip 0.4cm

Let us briefly sketch Hitchin's argument for $G=SL_2$.
There is only one (up to normalization) non-trivial 
invariant polynomial $P_2$ on $sl_2$ given by, say, 
half of the Killing
form. $\CH\equiv\CH_{P_2}$ maps into the space of quadratic 
differentials. A non-trivial holomorphic quadratic differential 
$\rho$ determines a (spectral) curve $\Sigma'\subset K$ given 
by the equation
\qq
\xi^2=\rho(\pi(\xi))
\label{1}
\qqq
where \s$\xi\in K$ and $\pi$ is the projection of $K$
on $\Sigma$. The map $\xi\mapsto-\xi$ gives an involution 
$\sigma$ of $\Sigma'$. Restriction of $\pi$ to $\Sigma'$ 
is a 2-fold covering of $\Sigma$ ramified over $4(\gamma-1)$       
points fixed by $\sigma$, the zeros of $\rho$. 
$\Sigma'$ has genus $\gamma'=4\gamma-3$. \m If \m$\rho\m=\m 
{_1\over^2}\s tr\s\s(\Phi)^2\m$ then 
relation (\ref{1}) coincides with the eigen-value equation
\qq
det\s(\Phi-\s\xi\cdot I)\s=\s 0
\nonumber
\qqq
for the Lax matrix $\Phi$.
Let for each $0\not=\xi\in\Sigma'$, $l_\xi$ denote
the corresponding eigen-subspace of $\Phi$. By continuity,
$l_\xi$ extend to vanishing $\xi$ in $\Sigma'$ and
\s$\cup_{_{\xi}}l_{\xi}$ forms a line subbundle $l$ 
of $\Sigma'\times\NC^2$. In fact, $l$ is a holomorphic subbundle
with respect to the complex structure defined on
$\Sigma'\times\NC^2$ by $\de+A\circ\pi$. The degree
of $l$ is $-2(\gamma-1)$. Besides,
\qq
l\s(\sigma^*l)\s=\s\pi^*K^{-1}\s.
\label{K}
\qqq    
Conversely, given $\Sigma'$ and a holomorphic 
line bundle $l$ of degree $-2(\gamma-1)$ on it satisfying 
(\ref{K}), we may recover a rank 2 holomorphic bundle $E$ 
of trivial determinant over $\Sigma$ as a pushdown of $l$ 
to $\Sigma$. Thus for $0\not=\xi\in\Sigma'$, $E_{\pi(\xi)}
=l_\xi\oplus l_{-\xi}$. $E$ corresponds to a unique
holomorphic $SL_2$-bundle which, if stable (what happens 
on an open subset of $l$'s) defines a point in the moduli 
space $\CN_s$. A holomorphic 1,0-form 
with values in the traceless endomorphisms of $E$
acting as multiplication by $\pm\xi$ on 
$l_{\pm\xi}\subset E_{\pi(\xi)}$ defines then
a unique covector of $T^*\CN_s$. Thus $\Sigma'$ encodes
the values of the quadratic Hitchin Hamiltonian $\CH$
(i.e.\s\s of the action variables) whereas the line bundles 
$l$ satisfying relation (\ref{K}) form the abelian (Prym) 
variety (of the angle variables) describing the level set 
of $\CH$.
\vs 0.9cm

\nsection{$SL_2$ moduli space at genus 2}
\vs 0.5cm

We shall present briefly the description
of the moduli space $\CN_s$ for $G=SL_2$
and $\gamma=2$ which was worked out in \cite{NarRa}.
\vskip 0.4cm

Let us start by recalling some basic facts about 
theta functions. We shall use a coordinate rather
than an abstract language. The space of degree $\gamma-1$ 
holomorphic line bundles 
forms a Jacobian torus $J^{\gamma-1}$ of complex dimension
$\gamma$. Fixing a marking (a symplectic homology basis
$(A_a,B_b),\ {a},{b}=1,\dots,\gamma$),
we may identify $J^{\gamma-1}$ with $\NC^\gamma/(\NZ^\gamma
+\tau\NZ^\gamma)$. $\tau\equiv(\tau^{{a}{b}})$ 
is the period matrix,
i.e. $\tau^{{a}{b}}=\int_{B_{{b}}}\omega^{a}$
where $\omega^{a}$ are the basic holomorphic
forms on $\Sigma$ normalized so that $\int_{A_{a}}
\omega^{b}=\delta^{{a}{b}}$. The point $0\in\NC^\gamma$
corresponds in $J^{\gamma-1}$ to a (marking dependent) spin 
structure $S_0$, i.e. a degree 1 bundle such that 
$S_0^2=K$. $u\in\NC^\gamma$ describes the line bundle 
$V(u)S_0$ where $V(u)$ is the flat line bundle 
with the twists $\ee^{2\pi i u^{b}}$ along the $B_{b}$ 
cycles. The set of degree 1 bundles $l$ with non-trivial
holomorphic sections forms a divisor $\Theta$ of a holomorphic
line bundle $L_\Theta$ over $J^{\gamma-1}$. Holomorphic
sections of the $k$-th power ($k>0$) of $L_\Theta$ 
are called theta function of order $k$. With the 
use of a marking, they may be represented by holomorphic 
functions $u\mapsto\theta(u)$ on $\NC^2$ satisfying
\qq
\theta(u+p+\tau q)\s=\s\ee^{-\pi i k\s q\cdot\tau q-2\pi i
k\s q\cdot u}\ \theta(u)
\qqq
for $p,q\in\NZ^\gamma$. The functions
\qq
\theta_{k,e}(u)\s=\s\sum\limits_{n\in{\dl Z}^\gamma}
\ee^{\s\pi i k\s
(n+e/k)\cdot\tau(n+e/k)\s+\s 2\pi i k\s(n+e/k)\cdot u}
\qqq
where $e\in\NZ^\gamma/k\NZ^\gamma$ form a basis of
the theta functions of order $k$. Hence $dim\s H^0(L_\Theta^k)
=k^\gamma.$ In particular, the Riemann theta function
$\theta_{1,0}(u)\equiv\vartheta(u)$ represents
the unique (up to normalization) non-trivial holomorphic
section of $L_\Theta$. It vanishes on the set
$$\{\sum\limits_{i=1}^{\gamma-1}\smallint_{x_0}^{x_i}\omega-
\Delta\ \vert\ x_1\in\Sigma,\dots,x_{\gamma-1}\in\Sigma\}$$ 
representing the divisor $\Theta$. Here $\Delta\in\NC^\gamma$ 
denotes the ($x_0$-dependent) vector of Riemann constants.
All theta functions of order 1 and 2 are even functions
of $u$.
\vskip 0.4cm

For $\gamma=2$, the divisor $\Theta$ is formed by the bundles
$\CO(x)$ with divisors $x\in\Sigma$. \m$\CO(x)=
V(\int_{x_0}^x\omega-\Delta)S_0$. The pullback 
of the theta bundle $L_\Theta$ by means of the map
$x\mapsto\CO(x)$ is equivalent to the canonical bundle $K$.
The equivalence assigns 1,0-forms to functions representing
sections of the pullback of $L_\Theta\m$:
\qq
\epsilon^{{a}{b}}\m\da_{b}\vartheta(
\smallint_{x_0}^x\omega-\Delta)\quad\mapsto\quad
\omega^a(x)\s.
\label{ide}
\qqq
This is consistent since vanishing 
of $\vartheta(\int\limits_{x_0}^x\omega-\Delta)$ implies that
\qq
\da_{a}\vartheta(\smallint_{x_0}^x\omega-\Delta)\ \omega^{a}(x)=0\ .
\nonumber
\qqq
Hence any multivalued function on $\Sigma$ picking up a factor 
$\ee^{-\pi i\tau^{{a}{a}}-2\pi i(\int_{x_0}^x\omega^{a}
-\Delta^{a})}$ when $x$ goes around the $B_{a}$
cycle and univalued around the $A_{a}$ cycles may be 
identified with a 1,0-form on $\Sigma$.
\vskip 0.4cm

As already suggested by the discussion at the end of 
Sect.\s\s 2, for the $SL_2$ group it is more convenient to
use the language of holomorphic vector bundles (of rank 2 and
trivial determinant) than to work with principal
$SL_2$-bundles. Of course the first ones are
just associated to the second ones by the fundamental
representation of $SL_2$. Any stable rank 2
bundle $E$ with trivial determinant is an extension 
of a degree 1 line bundle $l$
(\cite{NarRa}, Lemmas 5.5 and 5.8), i.e. it appears
in an exact sequence of holomorphic vector bundles 
\qq
0\m\longrightarrow\m l^{-1}\smash{\mathop{\longrightarrow}^\sigma}
\m E\m \smash{\mathop{\longrightarrow}^\varpi}
\s l\m \longrightarrow\m 0\ .
\label{ES}
\qqq
The inequivalent extensions (\ref{ES}) are classified
by the cohomology classes in $H^1(l^{-2})$. 
This may be seen as follows.
Taking a section of $\varpi$, \m i.e.\s\s a smooth bundle 
homomorphism $s:\s l\rightarrow E$ such that $\varpi\circ s=id_l$,
we infer that $\varpi\s\de s=0$ and hence that $\de s=\sigma\s b$
for $b$ a 0,1-form with values in $Hom(l,l^{-1})=l^{-2}$,
\m i.e. $b\in\wedge^{01}(l^{-2})$. $b$ is determined up to 
$\de\varphi$ where $\varphi$ is a smooth section of $l^{-2}$, 
i.e. $\varphi\in\Gamma(l^{-2})$. The class $[b]$ 
in $\wedge^{01}(l^{-2})/\Gamma(l^{-2})\s\cong\s
H^1(l^{-2})$ determines the extension (\ref{ES}) up
to equivalence. Each $b$ corresponds to an extension:
one may simply take $E$ equal to $l^{-1}\oplus l$
with the $\de$-operator given by $\de_{_{l^{-1}\oplus\m l}}+
\s(\matrix{_0&_b\cr^0&^0})$. \m Proportional $[b]$
correspond to equivalent bundles $E$. If $E$ is a stable
bundle then the extension (\ref{ES}) is necessarily
nontrivial, i.e. $[b]\not=0$.
\vskip 0.4cm

Let $C_E$ denote the set of degree 1 line 
bundles $l$ s.t. \s$H^0(l\otimes E)\not=0\s$
(equivalently, s.t. $E$ is an extension of $l$).
This is a complex 1-dimensional variety. It was shown
in \cite{NarRa} that $C_E$ characterizes the bundle
$E$ up to isomorphism and that there exists a
theta function $\theta$ of the 2$^{\rm nd}$ order which
vanishes exactly on $C_E$. The assignment
\s$E\s\mapsto\s\NC^*\theta\s$ gives an injective map 
\qq
m:\s\CN_s\s\longrightarrow\s\NP H^0(L_\Theta^2)\ .
\label{map}
\qqq
Let $V(u_1)S_0\equiv l_{u_1}\in C_E$. $E$ may be
realized as an extension of $l_{u_1}$ which is characterized
by $[b]\in H^1(l_{u_1}^{-2})$. Then one may take
\qq
\theta(u)\ = \int_{_\Sigma}\s K(x;u_1,u)
\wedge b(x)\ .
\label{thetaE}
\qqq
where
\qq
K(x;u_1,u)\ =\ \vartheta(\smallint_{x_0}^x\omega
-u_1-u-\Delta)\ \vartheta(\smallint_{x_0}^x
\omega-u_1+u-\Delta)\cr
\cdot\left(\epsilon^{{a}{b}}\s\da_{b}\vartheta
(\smallint_{x_0}^x\omega-\Delta)\right)^{-1}
\omega^{a}(x)
\label{Kk}
\qqq 
(it does not depend on the choice of $a=1,2$). \m
Let us explain the above formulae. \s$K(x;u_1,u)\m$, 
\m in its dependence
on \s$x\m$, \m is a multivalued holomorphic 1,0-form.
More exactly, the function
\qq
x\ \mapsto\ \vartheta(\smallint_{x_0}^x\omega-u_1-u-\Delta
)
\label{s_2}
\qqq
is multivalued around the $B_{a}$-cycles picking up the factor
$$\ee^{-\pi i\tau^{{a}{a}}\s-\s2\pi i(\int_{x_0}^x
\omega^{a}-u_1^{a}-u^{a}-\Delta^{a})}$$
when \s$x\s$ goes around \s$B_{a}\s$ so that it describes
an element \s$s_2\in H^0(l_{u_1} l_{u})\s$
(non-vanishing if \s$u_1+u\s\not\in\s\NZ^2+\tau\NZ^2\m$).
\m Similarly,
$$x\ \mapsto\ \vartheta(\smallint_{x_0}^x\omega
-u_1+u-\Delta)\ 
\left(\epsilon^{{a}{b}}\m\da_{b}\vartheta
(\smallint_{x_0}^x\omega-\Delta)\right)^{-1}
\omega^{a}(x)$$
picks up the factor $$\ee^{\s2\pi i(u_1^{a}-u^{a})}$$
when $x$ goes around $B_{a}$ and describes a 
holomorphic 1,0-form $\chi$ with values in $l_{u_1}l_u^{-1}\s$
(non-vanishing if \s$u_1-u\s\not\in\s\NZ^2+\tau\NZ^2\s$). 
\s The product \s$s_2\chi=K(\m\cdot\s;u_1,u)\s$ 
is a holomorphic 1,0-form with values in \s$l_{u_1}^{\m2}\s$ 
and it may be paired with \s$b\in\wedge^{01}(l_{u_1}^{-2})\s$ via
the integral over \s$x\s$ on the r.h.s. of Eq.\s\s(\ref{thetaE}).
The integral is independent of the choice of the representative $b$ 
of the cohomology class $[b]$. In its dependence on $u$,
\m$K(x;u_1,u)$ is a theta 
function of the 2$^{\rm nd}$ order and so is $\theta(u)$. 
In Appendix 1 we check explicitly that $\theta$ given
by Eq.\s\s(\ref{thetaE}) possesses the required property.
\vskip 0.4cm

The product of the two shifted Riemann 
theta functions \s$\vartheta(u'-u)\s
\vartheta(u'+u)\s$ is a theta function
of the 2$^{\rm nd}$ order both in $u'$ and in $u$
(and it is invariant under the interchange 
$u'\leftrightarrow u$). Let $\iota$ denote the (marking 
dependent) linear isomorphism between the spaces 
$H^0(L_\Theta^2)^*$ and $H^0(L_\Theta^2)$ defined by
\qq
\iota(\phi)(u)\s=\s{\langle\m}\vartheta(\s\cdot\s-u)
\s\m\vartheta(\s\cdot\s+u)\s\m,
\m\s\s\phi{\m\rangle} 
\label{lini}
\qqq
An easy calculation shows that
\qq
\vartheta(u'-u)\s\s\vartheta(u'+u)
\ =\ \sum\limits_{e}\theta_{2,e}(u')\s\s\theta_{2,e}(u)\s.
\qqq
Hence $\iota$ interchanges the basis $(\theta_{2,e})$ 
of $H^0(L_\Theta^2)$ with the dual basis $(\theta_{2,e}^{\s*})$
of $H^0(L_\Theta^2)^*$. Denote by $\phi_{u}$ the linear form 
on $H^0(L_\Theta^2)$ that computes the value of the theta 
function at point $u\in\NC^2$. The Kummer quartic 
$\CK^*\subset H^0(L_\Theta^2)^*$, \s$\CK^*
=\{\m\NC^*\phi_{u'}\s\vert\s u'\in\NC^2\m\}\m$ 
is mapped by the isomorphism $\iota$ into a quartic 
\s$\CK\subset H^0(L_\Theta^2)\s$ of theta functions  
proportional to
\qq
u\ \mapsto\ \vartheta(u'-u)\s\s\vartheta(u'+u)
\nonumber
\qqq
for some $u'\in\NC^2$.
\vskip 0.4cm

One may define a projective action of $(\NZ/2\NZ)^4$ 
on $H^0(L_\Theta^2)$ by assigning to an element 
$(e,e')\in(\NZ/2\NZ)^4$, with $e,e'=(0,0),\s(1,0),
\s(0,1)$ or $(1,1)$, a linear transformation $U_{e,e'}$ s.t.
\qq
(U_{e,e'}\theta)(u)\ =\ \ee^{\m{1\over2}\m\pi i\s
e'\cdot\m\tau\m e'\s+\s 2\pi i\s e'\cdot\m u}\s\s
\theta(u+\hf(e+\tau e'))\s.
\label{disa}
\qqq
The relation \s$U_{e_1,e_1'}\s U_{e_2,e_2'}\ =\ 
(-1)^{e_1\cdot e_2'}\s U_{e_1+e_2,\m e_1'+e_2'}\m$
holds so that $U$ lifts to the Heisenberg group.
In the action on the basic theta functions,
\qq
U_{e_1,e_1'}\m\theta_{2,e}\ =\ (-1)^{e_1\cdot\m e} 
\s\s\theta_{2,\m e+e_1'}\s.
\label{acone}
\qqq
The marking-dependence of the isomorphism
$\iota$ of Eq.\s\s(\ref{lini}) is given by the action 
of $(\NZ/2\NZ)^4$. It is easy to check that this 
action preserves \s$\CK$ and that the transposed action 
of $(\NZ/2\NZ)^4$ preserves $\CK^*$. The $(\NZ/2\NZ)^4$ 
symmetry of the Kummer quartics allows to find easily 
their defining equation, see Appendix 3. 
\vskip 0.4cm

It was shown in \cite{NarRa} that the image of $\CN_s$ 
under the map (\ref{map}) contains all non-zero
theta functions of the 2$^{\rm nd}$ order except the
ones in the the Kummer quartic $\CK$. 
The latter correspond, however, to the (Seshadri equivalence 
classes of) semistable but not stable bundles so that the map 
$m$ extends to an isomorphism between $\CN_{ss}$ and 
$\NP H^0(L_\Theta^2)$ showing that $\CN_{ss}$ is a smooth 
projective variety. 
\vskip 0.9cm

\nsection{Cotangent bundle}
\vskip 0.5cm

Let us describe the cotangent space of
$\CN_s$ at point $E$. The covectors tangent
to $\CN_s$ at $E$ may be identified with holomorphic
1,0-forms $\Psi$ with values in the bundle of traceless
endomorphisms of $E$. We may assume that $E$ is an 
extension of a line bundle $l$ of degree 1 realized as
$l^{-1}\oplus l$ with $\de_{_E}=\de_{_{l^{-1}\oplus\m l}}
\hs{-0.08cm}+B$ where \s$B=(\matrix{_0&_b\cr^0&^0})\m$. 
\s Then 
\qq
\Psi\ =\ \left(\matrix{-\mu&\nu\cr\eta&\mu}\right)
\label{Lax}
\qqq
where $\mu\in\wedge^{10}$, $\nu\in\wedge^{10}(l^{-2})$,
$\eta\in\wedge^{10}(l^2)$ and 
\qq
\de_{_{l^2}}\eta=0\s,\hspace{1cm}\de\mu=-\eta\wedge b\s,
\hspace{1cm}\de_{_{l^{-2}}}\nu=2\mu\wedge b\ .
\label{rel}
\qqq
It is easy to relate the above description of covectors
tangent to $\CN_s$ to the one of Sect.\s\s2.
Let $\CU:\s l^{-1}\oplus l\rightarrow\Sigma\times\NC^2$ be
a smooth isomorphism of rank 2 bundles with trivial determinant.
Then $\CU\s\de_{_E}\CU^{-1}=\de+A$ for a certain $sl_2$-valued 0,1-form
$A$ and $\Phi=\CU\Psi\CU^{-1}$ 
satisfies $\de_{_{A}}\Phi=0$. The $\CG$ orbit 
of $(A,\Phi)$ is independent of the choice of $\CU$
and the quadratic Hitchin Hamiltonian takes value 
${1\over 2}\s tr\s(\Phi)^2$ on it. The latter expression 
is clearly equal to ${_1\over^2}\s tr\s(\Psi)^2
=\s\mu^2+\eta\s\nu$ which, as easily follows from relations 
(\ref{rel}), defines a holomorphic quadratic differential. Hence
\qq
\CH(E,\Psi)\s=\s \mu^2+\eta\s\nu\ .
\label{hit}
\qqq
We would like to express the latter using the theta function
description of $T^*\CN_{ss}=T^*\NP H^0(L_{\Theta}^2)$
where the covectors tangent to $\CN_{ss}$ at $\NC^*\theta$ 
are represented by linear forms $\phi$ on $H^0(L_{\Theta}^2)$ 
\m s.t. ${\langle\m}\theta,\m\phi{\m\rangle}=0\m$.

\vskip 0.4cm

Let $l=l_{u_1}\in C_E$, \m i.e. $\theta(u_1)=0$
for the theta function corresponding to $E$.
We shall assume that $l^2\not=K$ i.e.
that \s$2u_1\s\not\in\s\NZ^2+\tau\NZ^2$. 
\m An infinitesimal variation $\delta E$ of the bundle $E$ 
in $\CN_s$ may be achieved by changing 
\s$\de_{_E}=\de_{_{l^{-1}\oplus\m l}}+B\s$ with \s$B=
(\matrix{_0&_b\cr^0&^0})\s$ to 
\qq
\de_{_{l^{-1}\oplus\m l}}+\left(\matrix{\pi\m\delta 
u_1({\rm Im}\s\tau)^{-1}\bar\omega
&b+\delta b\cr0&-\pi\m\delta u_1({\rm Im}\s\tau)^{-1}\bar\omega}
\right)\ \equiv\ \de_{_E}+\s\delta B
\label{DB}
\qqq
(all other variations of $\de_{_E}$ may be obtained from (\ref{DB}) 
by infinitesimal gauge transformations). Clearly
\qq
{\langle\m}\delta E\s,\s\Psi{\m\rangle}\ =\ \int_{_\Sigma}
tr\s\s\Psi\wedge\delta B\ =\ -\s 2\pi
\delta u_1({\rm Im}\s\tau)^{-1}\int_{_\Sigma}\mu\wedge\bar
\omega\s+\s\int_{_\Sigma}\eta\wedge\delta b\ .
\label{25}
\qqq
Note that the line bundle $l_{u_1}$ with the $\de$-operator 
changed to $\de_{_{l_{u_1}}}\hspace{-0.2cm}
-\pi\m\delta u_1({\rm Im}\s\tau)^{-1}
\bar\omega$ is equivalent to $l_{u_1+\delta u_1}\equiv l'$ and the
equivalence is established by multiplication by the
multivalued function
$x\mapsto\ee^{\s 2\pi i\s\delta u_1({\rm Im}\s\tau)^{-1}
\int_{x_0}^x{\rm Im}\s\omega}$. Hence $l^{-1}\oplus l$
with the $\de$-operator given by Eq.\s\s(\ref{DB}) is equivalent to 
${l'}^{-1}\oplus l'$ with the $\de$-operator $\de_{_{{l'}^{-1}\oplus
\m l'}}+\s(\matrix{_0&_{b+\delta'b}\cr^0&^0})\s$ where
\s$\delta'b(x)\ =\ \delta b\s -\s 4\pi i\s\delta u_1({\rm Im}
\s\tau)^{-1}(\smallint\limits_{x_0}^x{\rm Im}\s\omega)\s b(x)\m.$
\s The last bundle corresponds by the relation (\ref{thetaE})
to the theta function  
\qq
(\theta+\delta\theta)(u)\ =\ \int_{_\Sigma}
K(x;\m u_1+\delta u_1,\m u)\wedge (b(x)+\delta'b(x))\ .
\nonumber
\qqq
Hence $\delta E$ is represented by the variation 
\qq
\delta\theta(u)\ =\ 
-\ 2\pi\m\delta u_1^{a}\s ({\rm Im}\s\tau)^{-1}_{\s{a}{b}}
\int_{_\Sigma}L^b(x;u_1,u)\wedge b(x)\s
+\s\int_{_\Sigma}K(x;u_1,u)\wedge
\delta b(x)
\label{nd}
\qqq
of the theta function, where
\qq
L^{a}(x;u_1,u)\ =\ K(x;u_1,u)\s
\smallint_{x_0}^x(\omega^{a}-\bar\omega^{a})
\s-\s{_1\over^{2\pi}}{\rm Im}\s\tau^{{a}{b}}
\s\da_{u_1^{b}}K(x;u_1,u)\ .
\label{gal}
\qqq
Note that as functions of $x$, \s$L^a(x;u_1,u)$
are 1,0-forms with values in $l_{u_1}^2$ 
(as are $K(x;u_1,u)$). They are not holomorphic:
\qq
\de_x\m L^{a}(x;u_1,u)\ 
=\ K(x;u_1,u)\wedge\bar\omega^{a}(x)\s.
\nonumber
\qqq
As functions of $u$, \m$L^{a}(x;u_1,u)$ are theta 
functions of the 2$^{\rm nd}$ order. 
\vskip 0.4cm

We would like to find an explicit form of the Lax matrix $\Psi$ 
representing the linear form $\phi$ on $H^0(L_{\Theta}^2)$
s.t. ${\langle\m}\theta,\m\phi{\m\rangle}=0$. We shall
achieve this goal partially, finding the entries $\eta$
and $\mu$ of the matrix (\ref{Lax}). The correspondence 
between $\Psi$ and $\phi$ is determined by the equality
\qq
{\langle\m}\delta E\s,\s\Psi{\m\rangle}\ 
=\ {\langle\m}\delta\theta\s,\s\phi{\m\rangle}
\nonumber
\qqq
Since the left hand side is given by Eq.\s\s(\ref{25})
and $\delta\theta$ by Eq.\s\s(\ref{nd}), we obtain
\qq
&&\hs{1cm}-\s 2\pi\m
\delta u_1({\rm Im}\s\tau)^{-1}\int_{_\Sigma}\mu\wedge\bar
\omega\s+\s\int_{_\Sigma}\eta\wedge\delta b\cr
&&=\ 
-\m2\pi\m\delta u_1^{a}\s ({\rm Im}\s\tau)^{-1}_{\s{a}{b}}
\int_{_\Sigma}\hspace{-0.2cm}{\langle\m}L^b(x;u_1,\s\cdot\s)\m,
\s\phi{\m\rangle}\wedge\m b(x)\s
+\s\int_{_\Sigma}\hspace{-0.2cm}
{\langle\m}K(x;u_1,\s\cdot\s)\m,\s\phi{\m\rangle}\wedge\m\delta b(x)\s.
\hspace{0.5cm}
\label{equll}
\qqq
Taking $\delta u_1=0$ we infer that
\qq
\eta(x)\ =\ {\langle\m}K(x;u_1,\s\cdot\s)\s,\s\phi{\m\rangle}
\label{eta}
\qqq
is the lower left entry of the matrix $\Psi$ corresponding
to the linear form $\phi$. 
\vskip 0.4cm

It is easy to find the entry $\mu$ of $\Psi$ 
representing the linear form $\phi_{u_1}$ (recall
that $\phi_{u_1}$ computes the value of a theta 
function in $H^0(L_\Theta^2)$ at point $u_1$). 
Since $K(x;u_1,u_1)=0\m$, it follows from Eq.\s\s(\ref{eta})
that $\eta=0$ in this case. Eq.\s\s(\ref{equll}) reduces 
then to
\qq
&&-\s 2\pi\m\delta u_1({\rm Im}\s\tau)^{-1}
\int_{_\Sigma}\mu\wedge\bar
\omega\s=\s\delta u_1^{a}\int_{_\Sigma}
\da_{u_1^{a}} K(x;u_1,u_1)\wedge b(x)\s\cr
&&=\s -\s\delta u_1^{a}\int_{_\Sigma}
\da_{u^{a}} K(x;u_1,u_1)\wedge b(x)\s
=\s-\s\delta u_1^{a}\s\s\da_{a}
\theta(u_1)\s.
\nonumber
\qqq
This fixes $\mu$ uniquely:
\qq
\mu\ =\ {_i\over^{4\pi}}\s\da_{a}\theta(u_1)\s\s
\omega^{a}\ .
\label{mu}
\qqq
Let us check that there exists $\nu\in\wedge^{10}(l_{u_1}^{-2})$
such that the last equation of (\ref{rel}) holds. For this
it is necessary and sufficient that 
\qq
\int_{_\Sigma}{\kappa}\m\s\mu\wedge b\ =\ 0
\label{cons}
\qqq
for a non-zero holomorphic section ${\kappa}$ of \m$l_{u_1}^2=
V(2u_1)K$
(\m$dim\s H^0(l_{u_1}^2)=1$ if \s$2u_1\s\not\in\s\NZ^2+\tau\NZ^2$).
\m But such a section may be represented by the function
\qq
x\ \mapsto \ \vartheta(\smallint_{x_0}^x\omega-2u_1-\Delta)
\nonumber
\qqq 
so that, recalling the definition (\ref{Kk}), we obtain
\qq
\int_{_\Sigma}{\kappa}\m\s\omega^{a}\wedge b\ =\ 
\int_{_\Sigma}
\epsilon^{{a}{b}}\s\da_{u^{b}} K(x;u_1,u_1)\wedge b(x)
\ =\ \epsilon^{{a}{b}}\s\s\da_{b}\theta(u_1)\ .
\label{kob}
\qqq
Hence the relation (\ref{cons}) follows for $\mu$ given by
Eq.\s\s(\ref{mu}). The 1,0-form $\nu$ satisfying the last relation 
of (\ref{rel}) is now unique since $H^0(l_{u_1}^{-2}K)=\{0\}$.
\vskip 0.4cm

We would like to find the entry $\mu$ 
of $\Psi$ corresponding to more general linear forms 
$\phi$ s.t. ${\langle\m}\theta,
\m\phi{\m\rangle}=0$. Recall that $\theta$
with $\theta(u_1)=0$ may be given by formula (\ref{thetaE}) 
with $b\in\wedge^{0,1}(l_{u_1}^{-2})$. Note that any
2$^{\rm nd}$-order theta function \s$\delta\theta$
vanishing at $u_1$ and not in the Kummer quartic $\CK$
may be written as
\qq
\delta\theta(u)\ =\ \int_{_\Sigma}K(x;u_1,u)\wedge
\delta b(x)
\label{dth}
\qqq
with $\delta b\in\wedge^{01}(l_{u_1}^{-2})$ since 
it corresponds to an extension of $l_{u_1}$. 
The space of $\delta\theta$ vanishing at $u_1$
is 3-dimensional, as well as the space
$H^1(l_{u_1}^{-2})$ of classes $[\delta b]$ 
and the assumption that $\delta\theta\not\in\CK$ 
is obviously superfluous. Set for a linear form $\psi$ on 
$H^0(L_{\Theta}^2)$,
\qq
\eta_\psi(x)\s=\ {\langle\m}K(x;u_1,\s\cdot\s)\m,\s\psi{\m\rangle}\ .
\label{etab}
\qqq
$\eta_\psi$ defines a holomorphic 1,0-form with values
in $l_{u_1}^2$. We have
\qq
{\langle\m}\delta\theta\m,\m\psi{\m\rangle}\ =\s\int_{_\Sigma}
\eta_\psi\wedge\delta b
\label{dta}
\qqq
for \s$\delta\theta\s$ given by Eq.\s\s(\ref{dth}). By 
dimensional count, the map $\psi\mapsto\eta_\psi$ is onto 
$H^0(l_{u_1}^2K)$ with the 1-dimensional kernel spanned 
by $\phi_{u_1}$. Specifying Eq.\s\s(\ref{dta}) to $\delta\theta
\propto\theta$, we obtain the relation
\qq
{\langle\m}\theta\m,\m\psi{\m\rangle}\ =\s\int_{_\Sigma}\eta_\psi
\wedge b
\label{dtE}
\qqq
which determines the class $[b]\in H^1(l_{u_1}^{-2})$ in terms
of $\theta$. On the other hand, taking $\psi=\phi$ 
in Eq.\s\s(\ref{etab}), we infer that $\eta=0\m$ if and only 
if $\phi$ is proportional to $\phi_{u_1}$, 
the case studied before. 
\vskip 0.4cm

If $\eta_\phi\not=0$ then $\mu$ depends on the choice
of the representative $b$ in the class $[b]\in 
H^1(l_{u_1}^{-2})$ characterizing $E$ as the extension
of $l_{u_1}$. Under the transformation $b\mapsto b+\de\varphi$
where $\varphi$ is a section of $l_{u_1}^{-2}$, 
\qq
\eta\mapsto\eta\m,\ \quad\mu\mapsto\mu+\varphi\m\eta\m,\ \quad
\nu\mapsto\nu-2\m \varphi\m\mu-\varphi^2\eta\s.
\nonumber
\qqq
The pairing of the theta functions $L^a(x;u_1,\s\cdot\s)$  
of Eq.\s\s(\ref{gal}) with the linear form $\phi$
gives two 1,0-forms with values in $l_{u_1}^2$:
\qq
\chi^{a}(x)\ =\ {\langle\m}L^{a}(x;\m\cdot\s,u_1)\m,
\s\phi{\m\rangle}\quad\ \ 
{\rm s.t.}\quad\ \de\chi^{a}\s
=\s\eta\wedge\bar\omega^{a}\s.
\label{cih}
\qqq
Specifying the equality (\ref{equll}) to the case
with $\delta b=0$, we infer the relation
\qq
\int_{_\Sigma}\mu\wedge\bar\omega^a\ =\ \int_{_\Sigma}\chi^a
\wedge b
\label{cfm}
\qqq
which, together with the equation 
\qq
\de\mu=-\eta\wedge b
\label{cfm1}
\qqq
determines $\mu$ completely. In Appendix 2, we show 
that $\mu$ fixed this way satisfies the relation 
$\int_{_\Sigma}{\kappa}\s\mu\wedge b=0$ and hence defines 
a unique 1,0-form $\nu$ with values in 
$l_{u_1}^{-2}$ s.t. $\de\nu=2\m\mu\wedge b$.
\vskip 0.9cm

\nsection{Hitchin Hamiltonians}
\vskip 0.5cm

From the relation (\ref{hit}) and the explicit form of
$\Psi$ corresponding to $\phi_{u_1}$ ($\eta$ vanishing, $\mu$
given by Eq.\s\s(\ref{mu})\m), 
one obtains
\qq
\CH(\m\theta\m,\s a_1\m\phi_{u_1})\ =\ -\m{_1
\over^{16\pi^2}}\s a_1^2\s\s
(\s \da_{a}\theta(u_1)\s\s\omega^{a}\s)^2\s.
\label{first}
\qqq
The right hand side is a quadratic differential.
Eq.\s\s(\ref{first}), whose projective
version was first obtained in \cite{VGP}, 
is consistent with the rescaling $\theta\mapsto 
t\s\theta$ and $\phi\mapsto t^{-1}\phi$ for $t\in\NC^*$.
It describes the value of the Hitchin map
$\CH$ on the special covectors, namely those 
represented by the pairs $(\theta\m,\s\phi)$ 
s.t. $\NC^*\phi$ is in the intersection $\CK^*_E$
of the Kummer quartic $\CK^*$ with the plane 
${\langle\m}\theta,\m\phi{\m\rangle}=0$.
The linear span of $\CK^*_E$ gives the whole cotangent
space $T^*_E\CN_{ss}$. Indeed, any theta function of
the 2$^{\rm nd}$ order $\delta\theta$ which vanishes 
on $C_E$ has to be proportional to $\theta$ and defines 
a zero vector in $T_E\CN_{ss}$.
$\CK^*_E$ is itself a quartic. Hence the restriction 
of the quadratic polynomial $\CH$ to six lines in $\CK^*_E$ 
in a general position determines $\CH$ completely.
\vskip 0.4cm

It is possible to find a more explicit description
of the values of $\CH$ away from $\CK^*_E$ and this is
the main aim of the rest of the present section.
Suppose then that the entry $\eta$ in $\Psi$
does not vanish. Let $x_i$, $i=1,\dots,4$, be its four zeros.
We shall assume that $\eta$ cannot be written
as $\kappa\m\omega$ for $\kappa\in H^0(l_{u_1}^2)$ and 
$\omega\in H^0(K)$. This is true for 
generic $\phi$. In this case, \s$\eta\m=\m 
a_2\m\eta_{\phi_{u_2}}\s$ for some $a_2\in\NC^*$ 
and for $u_2$ satisfying
\qq
u_1+u_2\s=\s\smallint_{x_0}^{x_1}\omega
+\smallint_{x_0}^{x_2}\omega-2\Delta\ \quad{\rm and}\ \quad
u_1-u_2\s=\s \smallint_{x_0}^{x_3}\omega
+\smallint_{x_0}^{x_4}\omega-2\Delta\s,
\label{upm}
\qqq
$u_1\pm u_2\s\not\in\s\NZ+\tau\NZ$.
\s Indeed, \m$\eta_{\phi_{u_2}}(x)\m$ 
is a holomorphic section of $l_{u_1}^2K$
represented by the multivalued function
\s$\vartheta(\smallint_{x_0}^x
\omega-u_1-u_2-\Delta)\s\m\vartheta(\smallint_{x_0}^x\omega
-u_1+u_2-\Delta)\s$ vanishing exactly
at $x_i$ and such a section is unique up to normalization.
We infer that in the action on the theta functions
of Eq.\s\s(\ref{dth}), the linear 
forms $\phi$ and $a_2\m\phi_{u_2}$
coincide. Since Eq.\s\s(\ref{dth}) gives all theta functions
vanishing at $u_1$, it follows that 
\qq
\phi\s =\s a_1\s\phi_{u_1}\s+\s a_2\s\phi_{u_2}
\label{alpha}
\qqq
for some $a_1\in\NC$. Let us stress that, to fix normalizations, 
$u_1$ and $u_2$ should be viewed as elements of $\NC^2$ 
with $x_i$ in relations (\ref{upm}) belonging to the covering space 
$\tilde\Sigma$ of $\Sigma$. The relation 
${\langle\m}\theta\m,\m\phi{\m\rangle}\s=0$ 
implies that $\theta(u_2)=0$. 
\vskip 0.4cm

Summarizing, we have shown that a generic pair $(\theta,\phi)$ 
s.t. ${\langle\m}\theta,\m\phi{\m\rangle}=0$ 
may be obtained by first choosing $u_1$ and $u_2$ 
s.t. $2u_1,\m 2u_2,\m u_1\pm u_2\s\not\in\s\NZ+\tau\NZ$
\m and then taking $\theta$ from the 2-dimensional space of theta 
functions vanishing at $u_1$ and $u_2$ and $\phi$ from 
the orthogonal subspace. 
The zeros $x_i$ of $\eta$ are determined from Eqs.\s\s(\ref{upm})
(as the zeros of $\vartheta(\smallint_{x_0}^x
\omega-u_1\pm u_2-\Delta)$). For simplicity, we shall 
assume that they are distinct (this is true for generic $\phi$).
Then the differentials $\da\eta(x_i)\in(l_{u_1}^2K^2)_{x_i}$
do not vanish. 
\vskip 0.4cm

A quadratic differential $\rho\in H^0(K^2)$
is determined by its values at four points $x_i$ which form 
a divisor of $l_{u_1}^2K\not=K^2$. Since ${dim}\m H^0(K^2)=3$,
there is one linear relation satisfied by all $\rho(x_i)\m$:
\qq
\sum\limits_{i=1}^4\rho(x_i)\s{\kappa}(x_i)\s\da\eta(x_i)^{-1}\s=\s 0
\nonumber
\qqq
for $0\not={\kappa}\in H^0(l_{u_1}^2)$. \m It expresses the fact that 
the sum of residues of the meromorphic 1,0-form $\rho\m{\kappa}
\m\eta^{-1}$ has to vanish. For $\rho=\CH(\theta,\phi)=\mu^2+\eta
\nu$,
\qq
\rho(x_i)\s=\s\mu(x_i)^2
\nonumber
\qqq
so that it is enough to know $\mu(x_i)$ in order to determine
\m$\CH(\theta,\phi)$. \m Note that although
the 1,0-form $\mu$ depends on the choice of the representative $b$
of the class $[b]\in H^1(l_{u_1}^{-2})$ defined by Eq.\s\s(\ref{dtE}),
the values $\mu(x_i)$ are invariant since under $b\mapsto b+\de\varphi$
the 1,0-form $\mu$ changes to $\mu+\varphi\m\eta$.
\vskip 0.4cm

It remains to find $\mu(x_i)$. Consider the meromorphic function
$\eta_\psi\eta^{-1}$. Viewed as a distribution, 
$\de(\eta_\psi\eta^{-1})$ is supported at the poles 
of $\eta_\psi\eta^{-1}$ and
\qq
\int_{_\Sigma}\mu\wedge\de(\eta_\psi\eta^{-1})\s=\s
-2\pi i\sum\limits_{i=1}^4\mu(x_i)\m\eta_\psi(x_i)\s\da\eta(x_i)^{-1}
\nonumber
\qqq
for any (smooth) 1,0-form $\mu$. \m In particular, for $\mu$ 
satisfying Eq.\s\s(\ref{cfm1}) we obtain
\qq
\sum\limits_{i=1}^4\mu(x_i)\s\eta_\psi(x_i)\s\da\eta(x_i)^{-1}
\s=\s{_1\over^{2\pi i}}\int_{_\Sigma}\eta_\psi\wedge b\s
=\s{_1\over^{2\pi i}}\s\m{\langle\m}\theta,\m\psi{\m\rangle}\s.
\label{21}
\qqq
Recall that $\eta_\psi$ run through the three-dimensional
space $H^0(l_{u_1}^2K)$. If $\eta_\psi(x_i)=0$ for all $i$
then $\eta_\psi$ has to be proportional to 
$\eta=a_2\m\eta_{\phi_{u_2}}$. Hence vectors $(\eta_\psi(x_i))$ 
form a 2-dimensional subspace in $\mathop{\oplus}
\limits_i(l_{u_1}^2K)_{x_i}$ and equations (\ref{21})
determine vector $(\mu(x_i))\in\mathop{\oplus}\limits_i K_{x_i}$ 
up to a 2-dimensional ambiguity spanned by $(\omega^a(x_i))$
(indeed, as the residues of the meromorphic 1,0-form
$\eta_\psi\eta^{-1}\omega^a$, the numbers 
$\omega^a(x_i)\m\eta_\psi(x_i)\s\da\eta(x_i)^{-1}$ sum to zero).
It is clearly enough to take for $\psi$ in Eq.\s\s(\ref{21})
any two linear forms independent of $\phi_{u_1}$ and $\phi_{u_2}$.
In the generic situation, we may choose the forms 
$\da_a\phi_{u_1}$ defined by 
\qq
{\langle\m}\theta\m,\s\da_a\phi_{u_1}{\m\rangle}\s=\s\da_{a}\theta(u_1)\s.
\nonumber
\qqq
Denoting the corresponding 1,0-forms $\eta_\psi$ by $\eta'_a$,
we obtain 2 relations for $\mu(x_i)\m$:
\qq
\sum\limits_{i=1}^4\mu(x_i)\s\eta'_a(x_i)\s\da\eta(x_i)^{-1}
\s=\s{_1\over^{2\pi i}}\s\da_a\theta(u_1)\s.
\label{22}
\qqq
Alternatively, we may choose for $\psi$ the linear forms 
$\da_a\phi_{u_2}$ corresponding to 1,0-forms $\eta''_a$. 
This gives the relations
\qq
\sum\limits_{i=1}^4\mu(x_i)\s\eta''_a(x_i)\s\da\eta(x_i)^{-1}
\s=\s{_1\over^{2\pi i}}\s\da_a\theta(u_2)\s.
\label{24}
\qqq
$\eta''_a$ must be linearly dependent from $\eta'_a$ and $\eta$
(in the generic situation):
\qq
\eta''_a\ =\ D_a^b\s\eta'_b\s+\s\eta
\label{lind}
\qqq
leading via Eqs.\s\s(\ref{22}) and (\ref{24}) to the relation 
\qq
\da_a\theta(u_2)\ =\ D_a^b\s\s\da_b\theta(u_1)\s. 
\nonumber
\qqq
\vskip 0.4cm

We need 2 more equations to determine $\mu(x_i)$. They may be 
obtained from Eqs.\s\s(\ref{cfm}) fixing the holomorphic 
contributions to $\mu$. Indeed, using the 2$^{\rm nd}$ equation 
in (\ref{cih}), and Eq.\s\s(\ref{cfm1}) we infer that
\qq
\int_{_\Sigma}\mu\wedge\bar\omega^a\s=\s\int_{_\Sigma}(\mu\m\eta^{-1})
\s\eta\wedge\bar\omega^a\s=\s\int_{_\Sigma}(\mu\m\eta^{-1})\s\de\chi^a
\s=\s\int_{_\Sigma}\chi^a\wedge\de(\mu\m\eta^{-1})\cr
=\s\int_{_\Sigma}\chi^a\wedge b\s-\s2\pi i\sum\limits_{i=1}^4
\mu(x_i)\s\chi^a(x_i)\s\da\eta(x_i)^{-1}
\label{tbr}
\qqq
so that Eq.\s\s(\ref{cfm}) implies that
\qq
\sum\limits_{i=1}^4
\mu(x_i)\s\chi^a(x_i)\s\da\eta(x_i)^{-1}\s=\s0\s.
\label{mie}
\qqq
These are the two missing equations. To see this, 
repeat the calculation (\ref{tbr}) for $\mu$
replaced by $\omega^b$. This gives the relation
\qq
{_1\over^\pi}\s{\rm Im}\m\tau^{ab}\s=\s\sum\limits_{i=1}^4
\omega^b(x_i)\s\chi^a(x_i)\s\da\eta(x_i)^{-1}\s.
\nonumber
\qqq
Suppose now that \m$d_a\m\chi^a(x_i)+e\s\eta_\psi(x_i)=0$
\s for $i=1,\dots,4$. It follows that
\qq
0\s=\s\sum\limits_{i=1}^4
\omega^b(x_i)\left(d_a\chi^a(x_i)
+e\m\eta_\psi(x_i)\right)\da\eta(x_i)^{-1}
\s=\s {_1\over^\pi}\s{\rm Im}\m\tau^{ab}\s d_a
\nonumber
\qqq
so that $d_a=0$. Hence the vectors $(\chi^a(x_i))$ span
a 2-dimensional subspace of $\mathop{\oplus}\limits_iK_{x_i}$
transversal to the 2-dimensional subspace spanned
by the vectors $(\eta_\psi(x_i))$ and the linear equations
(\ref{21}) and (\ref{mie}) determine $\mu(x_i)$ completely.
\vskip 0.4cm

It is enough to consider the case $\phi=\phi_{u_2}$. 
\m Indeed, the shift \s$\phi\s\mapsto\s\phi\m+\m a_1
\m\phi_{u_1}\s$ results in the change
\qq
\mu\ \ \ \mapsto\ \ \ \mu\s+\s{_{i}\over^{4\pi}}\s a_1\s
\s\da_a\theta(u_1)\m\s\omega^a\s,
\nonumber
\qqq
see Eq.\s\s(\ref{mu}). Identifying 1,0-forms with 
multivalued functions by the relation (\ref{ide}) 
and setting \m$\chi_a=2\pi\m({\rm Im}\m\tau)^{-1}_{ab}
\chi^b$, \s${w}_i=\smallint_{x_0}^{x_i}\omega\m-\m\Delta$,
\s$G_1=G_{12}=-G_2$ and $G_3=G_{34}=-G_4$ where
\qq
G_{ij}\ =\ {det}\left(\matrix{\da_1\vartheta({w}_i)&
\da_1\vartheta({w}_j)\cr\da_2\vartheta({w}_i)&
\da_2\vartheta({w}_j)}\right)\s,
\nonumber
\qqq
we obtain
\qq
&&\da\eta(x_1)\s=\s G_1\ 
\vartheta({w}_1-{w}_3-{w}_4)\s,
\quad\quad\quad
\ \chi_a(x_1)\s =\s -\da_a\vartheta({w}_2)\ 
\vartheta({w}_1-{w}_3-{w}_4)\s,\cr
&&\da\eta(x_2)\s =\s G_2\ \vartheta({w}_2
-{w}_3-{w}_4)\s,
\quad\quad\quad
\ \chi_a(x_2)\s =\s -\da_a\vartheta({w}_1)\ 
\vartheta({w}_2-{w}_3-{w}_4)\s,\cr
&&\da\eta(x_3)\s =\s G_3\ 
\vartheta({w}_3-{w}_1-{w}_2)\s,
\quad\quad\quad
\ \chi_a(x_3)\s =\s -\da_a\vartheta({w}_4)\ 
\vartheta({w}_3-{w}_1-{w}_2)\s,\cr
&&\da\eta(x_4)\s =\s G_4\ 
\vartheta({w}_4-{w}_1-{w}_2)\s,
\quad\quad\quad
\ \chi_a(x_4)\s =\s -\da_a\vartheta({w}_3)\ 
\vartheta({w}_4-{w}_1-{w}_2)\s,\cr\cr
&&\eta'_a(x_1)\s =\s \da_a\vartheta({w}_1)\ 
\vartheta({w}_2+{w}_3+{w}_4)\s,\quad
\eta''_a(x_1)\s =\s\ \ \da_a\vartheta({w}_2)\ 
\vartheta({w}_1-{w}_3-{w}_4)\s,\cr
&&\eta'_a(x_2)\s =\s \da_a\vartheta({w}_2)\ 
\vartheta({w}_1+{w}_3+{w}_4)\s,\quad
\eta''_a(x_2)\s =\s\ \ \da_a\vartheta({w}_1)\ 
\vartheta({w}_2-{w}_3-{w}_4)\s,\cr
&&\eta'_a(x_3)\s =\s \da_a\vartheta({w}_3)\ 
\vartheta({w}_1+{w}_2+{w}_4)\s,\quad
\eta''_a(x_3)\s =\s -\da_a\vartheta({w}_4)\ 
\vartheta({w}_3-{w}_1-{w}_2)\s,\cr
&&\eta'_a(x_4)\s =\s \da_a\vartheta({w}_4)\ 
\vartheta({w}_1+{w}_2+{w}_3)\s,
\quad\eta''_a(x_4)\s =\s -\da_a\vartheta({w}_3)\ 
\vartheta({w}_4-{w}_1-{w}_2)\s.
\nonumber
\qqq
Given these values, it is easy to find the explicit
form of the matrix $(D^b_a)$ appearing 
in the relation between the derivatives
of $\da_a\theta$ at $u_1$ and $u_2$
by specifying Eq.\s\s(\ref{lind})
to two of the points $x_i$. One form of these relations is
\qq
&&\da_2\vartheta({w}_3)\s\da_1\theta(u_2)
\s-\s\da_1\vartheta({w}_3)\s\da_2\theta(u_2)\s\m\cr\cr
&&\hspace{2cm}=\s-\m{_{\vartheta({w}_3-{w}_1-{w}_2)}
\over^{\vartheta({w}_1+{w}_2+{w}_4)}}
\s\s(\da_2\vartheta({w}_4)\s\da_1\theta(u_1)
\s-\s\da_1\vartheta({w}_4)\s\da_2\theta(u_1))\s,\cr\cr
&&\da_2\vartheta({w}_4)\s\da_1\theta(u_2)
\s-\s\da_1\vartheta({w}_4)\s\da_2\theta(u_2)\s\cr\cr
&&\hspace{2cm}=\s-\m{_{\vartheta({w}_4-{w}_1-{w}_2)}
\over^{\vartheta({w}_1+{w}_2+{w}_3)}}
\s\s(\da_2\vartheta({w}_3)\s\da_1\theta(u_1)
\s-\s\da_1\vartheta({w}_3)\s\da_2\theta(u_1))\s.
\nonumber
\qqq
Let us denote
\s$\tilde\mu(x_i)\m=\m\mu(x_i)/G_i\m$. 
\m Eqs.\s\s(\ref{mie}) have the general solution
\qq
(\tilde\mu(x_1),\dots,\tilde\mu(x_4))\ =\ 
g_1\s\m(G_{34},0,G_{23},-G_{24})\s+\s
g_2\s\m(0,G_{34},G_{13},-G_{14})
\nonumber
\qqq
and Eqs.\s\s(\ref{24}) fix the values of $g_1$ and $g_2$ to 
\qq
&&g_1\ =\ -\m{\da_2\vartheta({w}_1)\s\da_1\theta(u_2)
\m-\m\da_1\vartheta({w}_1)\s\da_2\theta(u_2)\over
{4\pi i\s G_{12}\m G_{34}}}\cr\cr
&&g_2\ =\ \s\s\m{\da_2\vartheta({w}_2)\s\da_1\theta(u_2)
\m-\m\da_1\vartheta({w}_2)\s\da_2\theta(u_2)\over{4\pi i\s 
G_{12}\m G_{34}}}\s.
\nonumber
\qqq
This leads to the following simple result:
\qq
\mu(x_i)\s=\s\pm\s{_i\over^{4\pi}}\s\m(\m\da_2\vartheta({w}_i)
\s\da_1\theta(u_2)\s-\s\da_1\vartheta({w}_i)\s\da_2
\theta(u_2)\m)
\label{u1u2}
\qqq
or, in a more abstract notation from the introduction,
\qq
\mu(x_i)\s=\s\pm\s{_i\over^{4\pi}}\s\m d\theta(l_{u_2})
\nonumber
\qqq
with the plus sign for $i=1,2$ and the minus one for $i=3,4$.
\vskip 0.4cm

Since the Hitchin Hamiltonian is quadratic in $\phi$
and its values on $\phi_{u_1}$ and $\phi_{u_2}$
are given by Eq.\s\s(\ref{first}), it follows that
\qq
&&\CH(\m\theta\s,\m\s a_1\s\phi_{u_1}\m+\m a_2
\s\phi_{u_2})\cr\cr
&&\hspace{1cm}=\ a_1^2\s\CH(\m\theta\m,\s\phi_{u_1})\s
+\s a_2^2\s\CH(\theta\m,\s\phi_{u_2})
\s+\s2\m a_1a_2\s(\m c_1\m(\omega^1)^2\s
+\s c_2\m\omega^1\omega^2
\s+\s c_3\m(\omega^2)^2\m)\s.\hspace{0.8cm}
\nonumber
\qqq
The mixed term may be found from the linear equations
\qq
&&{_i\over^{4\pi}}\s(\da_2\vartheta({w}_i)\s\da_1\theta(u_1)-
\da_1\vartheta({w}_i)\s\da_2\theta(u_1))
\s\s\m\tilde\mu(x_i)\s\s G_i\cr\cr
&&\hspace{1.5cm}=\ c_1\s\da_2\vartheta
({w}_i)\s\da_2\vartheta({w}_i)
\s-\s c_2\s\da_2\vartheta({w}_i)\s\da_1\vartheta({w}_i)
\s+\s c_3\s\da_1\vartheta({w}_i)\s\da_1\vartheta({w}_i)\s.
\nonumber
\qqq
Their explicit solution leads to the expression
\qq
&&\hspace{-0.6cm}\CH(\m\theta\s,\m\s a_1\s\phi_{u_1}\m
+\m a_2\s\phi_{u_2})
\ =\ -\m{_1\over^{16\pi^2}}\s\s(\m a_1
\s\da_a\theta(u_1)\s\omega^a
\s+\s a_2\s\da_a\theta(u_2)\s\omega^a\m)^2\s\cr
&&+\s\s{_{a_1\m a_2}\over^{4\pi^2\s G_{13}\m G_{23}}}
\s\s(\da_2\vartheta({w}_3)\s\da_1\theta(u_1)
-\da_1\vartheta({w}_3)\s\da_2\theta(u_1))\s\clabel{ost}\cr
&&\ \ \ \cdot\ (\da_2\vartheta({w}_3)\s\da_1\theta(u_2)
-\da_1\vartheta({w}_3)\s\da_2\theta(u_2))
\ \da_a\vartheta({w}_1)\s\m\da_b
\vartheta({w}_2)\s\m\omega^a\omega^b\s.
\qqq
The second term on the right hand side hand side
is a quadratic differential that vanishes at $x_1$ and $x_2$
and is equal to \s${a_1\m a_2\over 4\pi^2}\s\s
\da_a\theta(u_1)\s\m\da_b\theta(u_2)\s\omega^a\omega^b\s$
at $x_3$ and $x_4$ so that
\qq
\CH(\theta,\m\phi)(x_i)\ =\ 
-\m{_1\over^{16\pi^2}}\s\s(\m a_1
\s\da_a\theta(u_1)\s\omega^a(x_i)
\s\pm\s a_2\s\da_a\theta(u_2)\s\omega^a(x_i)\m)^2\s
\label{frr}
\qqq
where sign plus should be taken for $x_1$ and $x_2$ and
sign minus for $x_3$ and $x_4$. This is the result
(\ref{fr}) described in Introduction.
\vskip 0.9cm

\nsection{Self-duality}
\vskip 0.5cm

We would like to compare the values of the Hitchin Hamiltonians
on the dual pairs $(\theta,\phi)$ and $(\theta',\m\phi')$ where
$\theta'=\iota(\phi)$ and $\phi'=\iota^{-1}(\theta)$
with $\iota$ defined by Eq.\s\s(\ref{lini}).
Recall that, given $u_1$ s.t. $\theta(u_1)=0$, we associated to 
the linear form \m$\phi\m$ a 1,0-form $\eta$
by Eq.\s\s(\ref{eta}).  Viewed as a holomorphic 
section of \s$l_{u_1}^2K$, 
\qq
\eta(x)\ =\ {\langle\m}\vartheta(\smallint_{x_0}^x\omega
-u_1-\s\cdot\s-\Delta)\ \vartheta(\smallint_{x_0}^x
\omega-u_1+\s\cdot\s-\Delta)\s\s,\ \phi{\m\rangle}\s.
\nonumber
\qqq
Let us denote
\qq
u_i'\s=\s\smallint_{x_0}^{x_i}\omega\m-\m u_1-\Delta\s.
\label{uip}
\qqq
The vanishing of $\eta(x_i)$ implies then that
the linear form $\phi$ annihilates the theta functions
\qq
u\ \ \mapsto\ \ \vartheta(u'_i-u)\s\m\vartheta
(u'_i+u)\s=\iota(\phi_{u'_i})(u)
\label{4theta}
\qqq
and also, if we rewrite $\eta(x_i)$ as $\iota(\phi)(u'_i)$,  
that $\theta'(u'_i)=0$. Since $\phi=
a_1\m\phi_{u_1}+a_2\m\phi_{u_2}$ and $\phi_{u_1}$ annihilates 
the theta functions (\ref{4theta}) as well, it follows that  
they belong to $\Pi$. Hence $\NC^*\iota(\phi_{u'_i})$ are 
the 4 points of intersection of the line $\NP\Pi$ with the Kummer 
quartic $\CK$. Equivalently, $\NC^*\phi_{u'_i}$ are the points
of intersection of $\NP\m{\Pi'}^\perp$ with $\CK^*$. 
In the generic situation, any pair of theta functions 
$\phi_{u'_i}$ spans ${\Pi'}^\perp$ and since 
$\phi'\in{\Pi'}^\perp$, we may write
\qq
\phi'\ =\ a'_1\s\phi_{v_1}\s+\s a'_2\s\phi_{v_2}
\label{aap}
\qqq
or, equivalently,
\qq
\theta\ =\ a'_1\s\iota(\phi_{v_1})\s
+\s a'_2\s\iota(\phi_{v_2})\s.
\label{aapp}
\qqq
\vskip 0.3cm

The involution $l\mapsto l^{-1}K$ of the Jacobian $J^1$ 
lifts to $\NC^2$ to the flip of sign of $u$. By restriction 
to the bundles $\CO(x)$, it induces the involution $x\mapsto x'$ 
of $\Sigma$ which leaves 6 Weierstrass points invariant.
The latter involution lifts to an involution
(without fixed points) of the covering space
$\tilde\Sigma$ determined by the equation
\qq
\smallint_{x_0}^{x}\omega\s-\s\Delta\ 
=\ -\smallint_{x_0}^{x'}\omega\s+\s\Delta\s.
\label{invo}
\qqq
Definitions (\ref{uip}) together with Eqs.\s\s(\ref{upm})
give the relations
\qq
u'_1-u'_2\s=\s\smallint_{x_0}^{x_1}\omega
-\smallint_{x_0}^{x_2}\omega\ \quad{\rm and}\ \quad
u'_1+u'_2\s=\s -\smallint_{x_0}^{x_3}\omega
-\smallint_{x_0}^{x_4}\omega+2\Delta
\nonumber
\qqq
holding in $\NC^2$, with $x_i\in\tilde\Sigma$.
They may be rewritten as
\qq
u'_1-u'_2\s=\s\smallint_{x_0}^{x_1}\omega
+\smallint_{x_0}^{x'_2}\omega\s-\s2\Delta\ \quad{\rm and}\ \quad
u'_1+u'_2\s=\s \smallint_{x_0}^{x'_3}\omega
+\smallint_{x_0}^{x'_4}\omega-2\Delta\s,
\label{upmq}
\qqq
which, upon the flip of the sign of $u'_2$ leaving 
$\phi_{u'_2}$ unchanged, provides the dual version
of relations (\ref{upm}) corresponding to points 
$x_1,x'_2,x'_3,x'_4\in\tilde\Sigma$.
Applying the previous result (\ref{frr}) and
using the possibility to exchange a point with
its image under the involution of $\Sigma$ in the argument
of a quadratic differential, we infer that
\qq
\CH(\theta',\m\phi')(x_i)\ =\ 
-\m{_1\over^{16\pi^2}}\s\s(\m a'_1
\s\da_a\theta'(u'_1)\s\omega^a(x_i)
\s\mp\s a_2\s\da_a\theta'(u'_2)\s\omega^a(x_i)\m)^2\s.
\label{frrr}
\qqq
The sign minus should be taken for $x_1$ and $x_2$ and sign plus 
for $x_3$ and $x_4$. The exchange of signs in comparison
with Eq.\s\s(\ref{frr}) is due to the flip $u'_2\mapsto-u'_2$.
\vskip 0.4cm

In order to compare expressions (\ref{frr}) and (\ref{frrr})
we shall calculate the coefficients $a_{1,2}$ and $a'_{1,2}$
of the linear combinations (\ref{alpha}) and (\ref{aap}).
Note that the definition $\theta'=\iota(\phi)$ implies that
\qq
\theta'(\smallint_{x_0}^x\omega-u_1-\Delta)\ =\ a_2\s\s\s
\vartheta(\smallint_{x_0}^x\omega-u_1-u_2-\Delta)
\s\s\vartheta(\smallint_{x_0}^x\omega-u_1+u_2-
\Delta)\s.
\nonumber
\qqq
Taking the derivative over $x$ at $x_1$, we obtain
\qq
\da_a\theta'(u'_1)\m\s\omega^a(x_1)\ = 
\ -\m a_2\s\s\s\vartheta(w_1-w_3-w_4)\s\s\da_a
\vartheta(w_2)\s\m\omega^a(x_1)
\nonumber
\qqq
where we employed Eqs.\s\s(\ref{upm}) and the abbreviated
notations $w_i=\smallint_{x_0}^{x_i}-\Delta$. Hence
\qq
a_2\ =\ -\m{_{\da_a\theta'(u'_1)\m\s\omega^a(x_1)}\over
^{\vartheta(w_1-w_3-w_4)\s\s\da_a
\vartheta(w_2)\s\m\omega^a(x_1)}}\s.
\label{a22}
\qqq
Similarly,
\qq
\theta'(\smallint_{x_0}^x\omega-u_2-\Delta)\ =\ a_1\s\s\s
\vartheta(\smallint_{x_0}^x\omega-u_1-u_2-\Delta)
\s\s\vartheta(\smallint_{x_0}^x\omega+u_1-u_2-
\Delta)\s.
\nonumber
\qqq
Taking the derivative at $x=x_1$ and noting that 
\m$w_1-u_2=-u'_2\m$, \m we infer that
\qq
a_1\ =\ {_{\da_a\theta'(u'_2)\m\s\omega^a(x_1)}\over
^{\vartheta(w_1+w_3+w_4)\s\s\da_a
\vartheta(w_2)\s\m\omega^a(x_1)}}\s.
\label{a11}
\qqq
To calculate $a'_{1,2}$, we note that Eq.\s\s(\ref{aapp})
implies that
\qq
\theta(\smallint_{x_0}^x\omega-v_1-\Delta)\ =\ a'_2\s\s\s\vartheta(
\smallint_{x_0}^x\omega-u'_1-u'_2-\Delta)\s\s
\vartheta(\smallint_{x_0}^x\omega-u'_1+u'_2-\Delta)\s.
\nonumber
\qqq
Upon derivation at $x=x_1$ and with the use of relations 
(\ref{upmq}) and (\ref{invo}), this gives
\qq
a'_2\ =\ -\m{_{\da_a\theta(u_1)\m\s\omega^a(x_1)}\over
^{\vartheta(w_1+w_3+w_4)\s\s\da_a
\vartheta(w_2)\s\m\omega^a(x_1)}}\s.
\label{a22p}
\qqq
Finally, since
\qq
\theta(\smallint_{x_0}^x\omega+v_2-\Delta)\ =\ a'_1\s\s\vartheta(
\smallint_{x_0}^x\omega-u'_1+u'_2-\Delta)\s\s
\vartheta(\smallint_{x_0}^x\omega+u'_1+u'_2-\Delta)\s.
\nonumber
\qqq
and \s$w_1+u'_2=u_2$ we infer that
\qq
a'_1\ =\ -\m{_{\da_a\theta(u_2)\m\s\omega^a(x_1)}\over
^{\vartheta(w_1-w_3-w_4)\s\s\da_a
\vartheta(w_2)\s\m\omega^a(x_1)}}\s.
\label{a11p}
\qqq
Substitution of expressions (\ref{a11}),(\ref{a22}),(\ref{a11p})
and (\ref{a22p}) shows equality of the right hand sides of
Eqs.\s\s(\ref{frr}) and (\ref{frrr}) for $x_i=x_1$. Since
there is a full symmetry between points $x_i$ (hidden in
our arbitrary choices of the order and the signs 
of $u_j\m$'s and $u'_j\m$'s), the self-duality  
\qq
\CH(\theta,\m\phi)\ =\ \CH(\theta',\m\phi')
\label{seld}
\qqq
follows.
\vskip 0.9cm

\nsection{van Geemen-Previato's result and beyond}
\vskip 0.5cm

The genus 2 curves are hyperelliptic.
The map \s$H^0(K)\ni\omega\s\mapsto\s\omega(x)\s$
defines an element of $\NP H^0(K)^*$ and varying $x\in\Sigma$
one obtains a realization of $\Sigma$ as a ramified double cover  
\m$\NP H^0(K)^*\cong\NP^1\m$. One may use the 1,0-forms 
$\omega^a\in H^0(K)$ to define the homogeneous coordinates 
on $\NP H^0(K)^*$. Then  
\qq
\lambda(x)\ =\ {_{\omega^2(x)}\over^{\omega^1(x)}}\ =\ 
-\m{_{\da_1\vartheta(\smallint_{x_0}^x\omega-\Delta)}
\over^{\da_2\vartheta(\smallint_{x_0}^x\omega-\Delta)}}
\label{inho}
\qqq
becomes the inhomogeneous coordinate of the image in $\NP^1$ 
of the point $x\in\Sigma$. \m If $x'$ is the image
of $x$ under the involution \s$\CO(x)\mapsto \CO(-x)K=\CO(x')\m$,
\s i.e.\s\s\m if
\qq
\smallint_{x_0}^x\omega+\smallint_{x_0}^{x'}\omega-2\Delta\ \in\ 
\NZ+\tau\NZ\quad\quad{\rm then}\quad\quad
\lambda(x)\s=\s\lambda(x')\s.
\nonumber
\qqq
Hence the involution $x\mapsto x'$ permutes the sheets of the 
covering $\Sigma\mapsto\NP^1$ ramified over the 6 Weierstrass 
points $x_s$, $s=1,\dots,6$, fixed by the involution. $\CO(x_s)$ 
is an odd spin structure. i.e. 
\qq
\smallint_{x_0}^{x_s}\omega-\Delta\ =\ E_s\ \ 
{\rm mod}\s(\NZ^2+\tau\NZ^2)
\nonumber
\qqq
and
\qq
\lambda_s\equiv\lambda(x_s)\s=\s
-\m{_{\da_1\vartheta(E_s)}
\over^{\da_2\vartheta(E_s)}}
\label{wp}
\qqq
where $E_s=\hf(e_s+\tau\m e_s')$ with $e_s,\m e_s'=(1,0),\s(0,1)$ 
or $(1,1)$ such that $e_s\cdot\m e_s'$ is odd. The possibilities 
are:
\qq
&&e_1=(1,0),\ e_1'=(1,0);\quad e_2=(1,1),\ e_2'=(1,0);
\quad e_3=(0,1),\ e_3'=(0,1);\cr\clabel{ordr}\cr
&&e_4=(1,1),\ e_4'=(0,1);\quad e_5=(0,1),\ e_5'=(1,1);
\quad e_6=(1,0),\ e_6'=(1,1).
\qqq
and we shall number the Weierstrass points (in a marking-dependent
way) in agreement with this list. $\Sigma$ may be identified 
with the hyperelliptic curve given by the equation 
\qq
\zeta^2\ =\ \prod\limits_{s=1}^6(\lambda-\lambda_s)
\label{hell}
\qqq
with the involution mapping $(\lambda,\zeta)$ to $(\lambda,-\zeta)$.
The expressions
\qq
\omega^1=C\s{{d\lambda}\over\zeta}\quad\ \ {\rm and}\quad\ \ 
\omega^2=C\s{{\lambda\m d\lambda}\over\zeta}\s,
\label{newf} 
\qqq
where $C$ is a constant, give the basis of holomorphic 1,0-forms 
of $\Sigma$ (the right hand sides vanish exactly where
the left hand sides do).
\vskip 0.4cm

Let us recall the main result of \cite{VGP} based on 
the analysis of the formula (\ref{first}) for the Hitchin 
Hamiltonians on the Kummer quartic $\CK^*$. It will be 
convenient to identify the pairs $(\theta,\phi)$ 
s.t. ${\langle\m}\theta,\phi{\m\rangle}=0$ 
with pairs $(q,p)\in\NC^4\times\NC^4$ s.t. $q\cdot p=0$
by the relations
\qq
&&\theta\ =\ q_1\s\theta_{2,(0,0)}\s+\s q_2\s\theta_{2,(1,0)}
\s+\s q_3\s\theta_{2,(0,1)}\s+\s q_4\s\theta_{2,(1,1)}\s,\cr
&&\phi\ =\ p_1\s\theta_{2,(0,0)}^{\s*}\s+\s p_2\s\theta_{2,(1,0)}
^{\s*}\s+\s p_3\s\theta_{2,(0,1)}^{\s*}\s+\s 
p_4\s\theta_{2,(1,1)}^{\s*}\s.
\nonumber
\qqq
The symplectic form of $T^*\NP^3$ is the standard
$dp\wedge dq$ and the isomorphism $\iota$ interchanges $p$
and $q$. By examining the values of the quadratic differentials 
given by $\CH$ at the Weierstrass points $x_s$, van Geemen and 
Previato showed that
\qq
\CZ_s(q)\ =\ \{\s p\s\s|\s\s q\cdot p=0,\ 
\CH(q,p)(x_s)=0\s\}
\nonumber
\qqq
is a union of a pair of bitangents to $\CK^*$.
Then classical results giving the equations
for bitangents to the Kummer surface permitted the authors
of \cite{VGP} to write an almost explicit formula for $\CH(x_s)$ 
in the form
\qq
\CH(q,p)(x_s)\ =\ h_s\s\sum\limits_{t\not= s}{{r_{st}(q,p)}
\over{\lambda_s-\lambda_t}}
\label{VGP}
\qqq
where $r_{st}=r_{ts}$ are homogeneous polynomials, 
\qq
&&r_{12}(q,p)\ =\ \ \ \m(q_1p_1+q_2p_2-q_3p_3-q_4p_4)^2\s,\cr
&&r_{13}(q,p)\ =\ \ \ \m(q_1p_4-q_2p_3-q_3p_2+q_4p_1)^2\s,\cr
&&r_{14}(q,p)\ =\ -(q_1p_4+q_2p_3-q_3p_2-q_4p_1)^2\s,\cr
&&r_{15}(q,p)\ =\ -(q_1p_3-q_2p_4-q_3p_1+q_4p_2)^2\s,\cr
&&r_{16}(q,p)\ =\ \ \ \m(q_1p_3+q_2p_4+q_3p_1+q_4p_2)^2\s,\cr
&&r_{23}(q,p)\ =\ -(q_1p_4-q_2p_3+q_3p_2-q_4p_1)^2\s,\cr
&&r_{24}(q,p)\ =\ \ \ \m(q_1p_4+q_2p_3+q_3p_2+q_4p_1)^2\s,\cr
&&r_{25}(q,p)\ =\ \ \ \m(q_1p_3-q_2p_4+q_3p_1-q_4p_2)^2\s,\clabel{rs}\cr
&&r_{26}(q,p)\ =\ -(q_1p_3+q_2p_4-q_3p_1-q_4p_2)^2\s,\cr
&&r_{34}(q,p)\ =\ \ \ \m(q_1p_1-q_2p_2+q_3p_3-q_4p_4)^2\s,\cr
&&r_{35}(q,p)\ =\ \ \ \m(q_1p_2+q_2p_1+q_3p_4+q_4p_3)^2\s,\cr
&&r_{36}(q,p)\ =\ -(q_1p_2-q_2p_1-q_3p_4+q_4p_3)^2\s,\cr
&&r_{45}(q,p)\ =\ -(q_1p_2-q_2p_1+q_3p_4-q_4p_3)^2\s,\cr
&&r_{46}(q,p)\ =\ \ \ \m(q_1p_2+q_2p_1-q_3p_4-q_4p_3)^2\s,\cr
&&r_{56}(q,p)\ =\ \ \ \m(q_1p_1-q_2p_2-q_3p_3+q_4p_4)^2\s
\qqq
and $h_s\in K^2_{x_s}$ could still depend on $q$.
In the original language of pairs $(\theta,\phi)$,
and of the $(\NZ/2\NZ)^4$-action (\ref{acone})
on $H^0(L_\Theta^2)$ one has
\qq
r_{st}(\theta,\phi)\ =\ \ {\langle\m}U_{e_s,e'_s}U_{e_t,e'_t}\m\theta\m,
\s\phi{\m\rangle}\s{\langle\m}U_{e_t,e'_t}U_{e_s,e'_s}
\m\theta\m,\s\phi{\m\rangle}
\nonumber
\qqq
with $e_s,\m e'_s$ from the list (\ref{ordr}).
The polynomials $r_{st}$ are self-dual:
\qq
r_{st}(q,p)\ =\ r_{st}(p,q)
\qqq
and the self-duality of $\CH$ proven in the present paper forces 
coefficients $h_s$ in Eq.\s\s(\ref{VGP}) to be $q$-independent 
filling partially the gap left in \cite{VGP}. 
An easy but important identity is
\qq
\sum\limits_{t\not=s}r_{st}(q,p)\ =\ (q\cdot p)^2\ =\ 0
\label{sum}
\qqq
for any fixed $s$. It implies that the Hamiltonians
(\ref{VGP}) are preserved up to normalization
by the isomorphisms of the hyperelliptic surfaces induced
by the fractional action \s$\lambda\s\mapsto\s\lambda'=
{a\lambda+b\over c\lambda+d}\s$ of $SL(2,\NC)$ on $\NP^1$.
\vskip 0.4cm

We would still like to fix the values of the constants
$h_s$ in Eqs.\s\s(\ref{VGP}). We claim that they are such 
that the Hitchin map is given by Eq.\s\s(\ref{GR}), i.e. that
\qq
\CH(q,p)\ \ =\ \ -\m{_1\over^{128\m\pi^2}}\s\m\sum\limits_{s,
t=1,\dots,6,\atop s\s\not=\s t}{r_{st}(q,p)\over
(\lambda-\lambda_s)(\lambda-\lambda_t)}\s\m(d\lambda)^2\s.
\label{glr}
\qqq
First note that the above formula is consistent with the
$SL(2,\NC)$ transformations. Indeed, relations
(\ref{sum}) imply that
\qq
\sum\limits_{s\s\not=\s t}{r_{st}\over
(\lambda'-\lambda'_s)(\lambda'-\lambda'_t)}\s\m(d\lambda')^2
\ =\ \sum\limits_{s\s\not=\s t}{r_{st}\over
(\lambda-\lambda_s)(\lambda-\lambda_t)}\s\m(d\lambda)^2
\nonumber
\qqq
for $\lambda'={a\lambda+b\over c\lambda+d}$. 
Taking, in particular, $\lambda'=\lambda^{-1}$ one verifies
that the quadratic differentials (\ref{glr}) are
regular at infinity. They are also regular at the branching
points since ${d\lambda\over\sqrt{\lambda-\lambda_s}}$
is a local holomorphic differential around $x_s$. 
Hence the r.h.s. of Eq.\s\s(\ref{glr}) is indeed
a (holomorphic) quadratic differential. Thus 
Eq.\s\s(\ref{glr}) is equivalent to relations (\ref{VGP})
with \s$h_s= {(d\lambda)^2\over(\lambda-\lambda_s)}
\vert_{_{x_s}}\m$, \s modulo an overall normalization. 
To prove Eq.\s\s(\ref{glr}) we shall verify it 
at a point of the phase space for which $\CH(q,p)(x_s)\not=0$
for $s\not=1$. This will fix $h_s$ for $s\not=1$ and hence
all of them (two quadratic differentials equal at points
$x_s$ with $s\not=1$ have to coincide).
\vskip 0.4cm

Consider a pair $(\theta,\phi_{u_1})$ lying in the product
$\CK\times\CK^*$ of the Kummer quartics with
\qq
\theta(u)\ =\ \ee^{{1\over 2}\m\pi i\s e'_1\cdot\tau\m e'_1\m
+\m 2\pi i\s e'_1\cdot u_1}\ \vartheta(u_1+E_1-u)
\ \vartheta(u_1+E_1+u)\hspace{1.5cm}\cr\cr
=\s\sum\limits_e\s(U_{e_1,e'_1}\theta_{2,e})(u_1)\ 
\theta_{2,e}(u)
\label{121}
\qqq
for $e_1=e'_1=(1,0)$. Note that ${\langle\m}\theta,\phi_{u_1}
\m\rangle=0$.
Eq.\s\s(\ref{first}) together with the relations (\ref{newf}) 
and the equation
\qq
\da_a\theta(u_1)\ =\ -\m\ee^{{1\over 2}\m\pi i\s e_1'\cdot\m\tau 
\m e'_1\s+\s 2\pi i\s e'_1\cdot\m u_1}\s\s\m\da_a\vartheta(E_1)
\ \vartheta(2u_1+E_1)
\nonumber
\qqq
results in the identity
\qq
\CH(\theta,\phi_{u_1})\ =\  
-\m {_{C^2}\over^{16\pi^2}}\s\s\ee^{\m\pi i\s e'_1\cdot\m\tau\m e'_1
\s+\s 4\pi i\s e'_1\cdot\m u_1}\s\s(\da_2\vartheta(E_1))^2\s\s 
\vartheta(2u_1+E_1)^2\s\s(\lambda-\lambda_1)^2
\s\s{_{(d\lambda)^2}\over^{\zeta^2}}
\hspace{0.4cm}
\label{123}
\qqq
where $C$ is the constant appearing in Eq.\s\s(\ref{newf}).
Note that $\CH(\theta,\phi_{u_1})\not=0$ as long as
\s$\vartheta(2u_1+E_1)\not=0$. It follows 
that $\CH(\theta,\phi_{u_1})$ is a quadratic
differential proportional to \s$(\lambda-\lambda_1)^2  
\s{(d\lambda)^2\over\zeta^2}$ which has the $4^{\m\rm th}$ 
order zero at $x_1$. The latter property characterizes it
uniquely up to normalization.
\vskip 0.4cm

It is not difficult to check that Eq.\s\s(\ref{glr}) gives 
a quadratic differential with the same property. Indeed,
in the language of $q\m$'s and $p\m$'s, the linear form 
\m$\phi_{u_1}$ corresponds to a vector $p\in\NC^4$ 
and $\theta$ to $q=(p_2,-p_1,p_4,-p_3)$. A straightforward
verification shows that $r_{1\m t}(q,p)=0$ for all $t\not=1$.
This implies that the quadratic differential given
by Eq.\s\s(\ref{glr}) vanishes to the second order
at $x_1$. The condition that it vanishes to the fourth 
order is
\qq
\sum\limits_{s\s\not=\s t,\atop s,t\s\not=\s 1}
r_{st}((p_2,-p_1,p_4,-p_3),\s p\m)\ 
\prod\limits_{v\s\not=\s1,s,t}
(\lambda_1-\lambda_v)\ =\ 0\s.
\nonumber
\qqq
A direct calculation shows that this is exactly the equation
(\ref{kuma}) of the Kummer quartic with the coefficients
(\ref{kumb}) so that it holds for
$p$ corresponding to $\phi_{u_1}$. This establishes
proportionality between the Hitchin map and
the right hand side of Eq.\s\s(\ref{glr}) with 
a coefficient that may be still curve-dependent. 
\vskip 0.4cm

Fixing the overall normalization of the Hitchin map
is more involved. We shall calculate the value
of the quadratic differential on the right hand side 
of Eq.\s\s(\ref{123}) at $\lambda=\lambda_2$ and 
compare it to the value given by Eq.\s\s(\ref{glr}). 
Since this is somewhat technical, we defer the argument
to Appendix 4.
\vskip 0.4cm

The system with Hamiltonians (\ref{VGP}) bears some similarity 
to the classic Neumann systems\footnote{we thank M. Olshanetsky 
for attracting our attention to this fact}, also anchored
in modular geometry \cite{Mum}\cite{AT}. The Hamiltonians
of a Neumann system have the form
\qq
\CH_s\ =\ \sum\limits_{1\leq t\not= s\leq n}{J_{st}^2
\over{\lambda_s-\lambda_t}}
\label{Neum}
\qqq
where $J_{st}\s=\s q_sp_t-q_tp_s$ are the functions on $T^*\NC^n$
generating the infinitesimal action of the complex group $SO_n$:
\qq
&&\hbox to 5cm{$\{J_{st},J_{tv}\}\s=\s -J_{sv}$\hfill}{\rm for\ \ }
s,t,v\ \ {\rm different},\cr
&&\clabel{son}\cr
&&\hbox to 5cm{$\{J_{st},J_{vw}\}\s=\s0$\hfill}{\rm for\ \ }
s,t,v,w\ \ {\rm different}.
\qqq
The fact that the Hamiltonians (\ref{VGP}) (with constant $h_s$)
Poisson commute reduces, as is well known, to the identities
\qq
&&\hbox to 5cm{$\{r_{st}+r_{sv}\m,\s r_{tv}\}\s=\s 0$\hfill}
{\rm and\  cyclic\ permutations\ thereof}\s,\cr
\clabel{cyb1}\cr
&&\hbox to 5cm{$\{r_{st}\m,\s r_{vw}\}\s=\s0$\hfill}
{\rm for}\quad\quad\{s,t\}\cap\{v,w\}=\emptyset\s.
\qqq
If we set $r_{st}=J_{st}^{\m2}$ for the Neumann system, then 
Eqs.\s\s(\ref{cyb1}) follow from the relations (\ref{son}). 
It appears that the same algebra stands behind 
the fact\footnote{this is the classical version 
of the observation of \cite{VGDJ}}
that $r_{st}$ given by Eq.\s\s(\ref{rs}) verify 
(\ref{cyb1}). The phase space \s$T^*\CN_{ss}\s\cong\s
\{\s(q,p)\s\m\vert\s\m q\cdot p=0\s\}/\NC^*\m$, \m where
$\NC^*$ acts by $(q,p)\mapsto(tq,\m t^{-1}p)$, may be 
identified with the coadjoint orbit of the group $SL_4$ 
composed of the traceless complex 4$\times$4 matrices 
$\vert p\rangle\langle q\vert$ of rank 1. Using the isomorphism 
of the complex Lie algebras $sl_4\cong so_6$, we obtain 
the functions $J_{st}=-J_{ts}$ on this $SL_4$-orbit  
which generate the action of $so_6$ and have the Poisson
brackets given by (\ref{son}). A straightforward check shows
that, for $r_{st}$ of Eq.\s\s(\ref{rs}),
\qq
r_{st}\ =\ -\m4\m J_{st}^{\m 2}
\qqq
so that Eq.\s\s(\ref{cyb1}) follows from the $so_6$-algebra
(\ref{son}). 
\vskip 0.4cm

Upon the introduction of the rational 
functions \s${r_{st}\over\lambda}\m$,
Eqs.\s\s(\ref{cyb1}) take the form 
\qq
&&\{{_{r_{st}}\over^{\lambda_s-\lambda_t}}\m,\s
{_{r_{sv}}\over^{\lambda_s-\lambda_v}}\}\ +\ \{
{_{r_{st}}\over^{\lambda_s-\lambda_t}}\m,\s
{_{r_{tv}}\over^{\lambda_t-\lambda_v}}\}
\ +\ \{{_{r_{sv}}\over^{\lambda_s-\lambda_v}}\m,
\s{_{r_{tv}}\over^{\lambda_t-\lambda_v}}\} 
\ =\ 0\s,\cr
\clabel{cyb2}\cr
&&\{{_{r_{st}}\over^{\lambda_s-\lambda_t}}\m,\s
{_{r_{vw}}\over^{\lambda_v-\lambda_w}}\}\ =\ 0\ 
\quad\quad\quad {\rm for}\quad\ \{s,t\}
\cap\{v,w\}=\emptyset\s.
\qqq
The first of these identities is, essentially, the classical 
Yang-Baxter equation. Note, however, that $r_{st}$, 
unlike in the Gaudin and Neumann systems, is not an element 
of a product of two copies of a Poisson algebra 
of functions: there is no sign of an explicit product 
structure, or of a reduction thereof, in our phase space. 
The important question is whether $r_{st}$ come from 
a rational solution of the CYBE. The conformal field theory 
work \cite{Knizh}\cite{Zamo} suggests that the answer may 
be positive, at least in some sense. 
\vskip 0.4cm

The knowledge of the explicit form of the quadratic differentials
$\CH(q,p)$ allows to write the explicit equations for
the genus 5 spectral curve of the $SL_2$ Hitchin system at genus 2,
see Eq.\s\s(\ref{1}). They take the form
\qq
\zeta^2\ =\ \prod\limits_{s=1}^6(\lambda-\lambda_s)\s,\quad\ \ 
\xi^2\ =\ \sum\limits_{s\s\not=\s t}r_{st}(q,p)
\prod\limits_{v\s\not=\s s,t}(\lambda-\lambda_v)\s.
\label{specc}
\qqq
The involution of the spectral curve flips the sign of $\xi$.
To extract explicit formulae for the angle variables describing 
the point on the Prym variety of the spectral curve, we would
need, however, a more explicit knowledge of the entire
Lax matrix $\Psi$.
\vskip 0.9cm

\nsection{Conclusions}
\vskip 0.5cm

The main result of the present paper is the proof
of self-duality of the Hitchin Hamiltonians 
on the cotangent bundle to the moduli space 
of the holomorphic $SL_2$ bundles on a genus 2 complex 
curve. The result was based on an expression for 
the Hitchin Hamiltonians off the Kummer quartic on
which the values of the Hamiltonians were determined
in \cite{VGP}. Using the self-duality,
we were able to complete the analysis of \cite{VGP}
and to obtain the explicit formula (\ref{GR})
for the Hitchin map (\ref{Hm}) giving the action
variables of the integrable system. The explicit 
formula for the angle variables remains still to be found. 
An interesting open problem is an extension of the present 
work to the case with insertion points.
\vskip 0.4cm

Another important problem related to Hitchin's construction 
is the quantization of the corresponding integrable systems. 
For the $SL_2$ case such a quantization is essentially
provided by the Knizhnik-Zamolodchikov-Bernard-Hitchin
connection \cite{KZ}\cite{Ber}\cite{Ber2} which describes 
the variation of conformal blocks of the $SU_2$ WZW conformal
field theory under the change of the complex structure 
of the curve. The (partition function) conformal blocks 
are holomorphic sections of the $k^{\m\rm th}$-power 
of the determinant line bundle over the moduli space $\CN_{ss}$ 
($k$ is the level of the WZW theory). In our case, they are simply 
$k^{\m\rm th}$-order homogeneous polynomials on $H^0(L_\theta^2)$.
It is easy to quantize the Hitchin Hamiltonians
\qq
H_s\ =\ \sum\limits_{t\s\not=\s s}{r_{st}\over\lambda_s-\lambda_t}\s.
\nonumber
\qqq
If one keeps the original formulae (\ref{rs}) for $r_{st}$ 
in which $p_i$ stands now for ${1\over i}\s\da_{q_i}$, the
relations (\ref{cyb1}) or (\ref{cyb2}) still hold after
the replacement of the Poisson brackets by the commutators. 
One obtains this way the commuting operators $H_s$ mapping 
the space of homogeneous, degree $k$ polynomials 
in variables $q$ into itself. Note, however, that now
\qq
\sum\limits_{t\s\not=\s s}r_{st}\ =\ -\m k(k+4)
\nonumber
\qqq
for each fixed $s$ so that the quantization changes the conformal 
properties of the Hamiltonians. A direct construction of 
the projective version of the KZBH connection for group $SU_2$ 
and genus 2 has been recently given in ref.\s\s\cite{VGDJ} 
by following Hitchin's approach \cite{Hitch2}. It is consistent
with the above {\it ad hoc} quantization of the classical
Hitchin Hamiltonians.
\vskip 0.4cm

The integral formulae for the conformal blocks 
\cite{BabF,ReshV,EtinK} or, equivalently, 
the integral formulae for the scalar product of
the conformal blocks \cite{FalKG} have been used
at genus 0 and 1 to extract the Bethe Ansatz
eigen-vectors and eigen-values of the quantized 
version of the quadratic Hitchin Hamiltonians.
The Bethe-Ansatz type diagonalization of the 
quantization of the genus 2 Hitchin Hamiltonians
is among the issues that will have to be examined.
\vskip 0.4cm

Finally, as we stressed in the text, the relations
between the conformal WZW field theory on a genus 2
surface and an orbifold theory in genus 0 requires 
further study.
\vskip 0.9cm

\renewcommand{\theequation}{A\thesection.\arabic{equation}}
\def\theequation{{A\thesection.\arabic{equation}}}
\nappendix{1}

\vskip 0.5cm

Let us check that $\theta$ given by Eq.\s\s(\ref{thetaE})
vanishes if and only if $$H^0(l_u\otimes E)\s
=\s\{\s(s_1,s_2)\ |\ s_2\in H^0(l_u 
l_{u_1}),\ \de_{_{l_u^{-1}l_{u_1}}}
\hspace{-0.1cm}s_1+s_2\s b=0\s\}\s\ \not=\ \s0\s.$$
For \s$u-u_1\in\NZ^2+\tau\NZ^2\s$ the 1$^{\rm st}$
theta function on the r.h.s. of Eq.\s\s(\ref{Kk}) vanishes
but  \s$l_u=l_{u_1}\s$ and $l_{u_1}\in C_E\m$. Assume
now that \s$u-u_1\s\not\in\s\NZ^2+\tau\NZ^2\m$.
Then \s${dim}\ H^0(l_u^{-1}l_{u_1}K)=1\s$
with a non-zero \s$\chi\in H^0(l_u^{-1}l_{u_1}K)$. 
The necessary and sufficient condition for the solvability 
of the equation \s$\de_{_{l_ul_{u_1}^{-1}}}s_1+s_2\m b=0\s$ 
for a given \s$s_2\in H^0(l_ul_{u_1})\s$ is 
\qq
\int_{_\Sigma}\chi\m s_2\s b=0\s.
\label{soco}
\qqq
If \s$u+u_1\in \NZ^2+\tau\NZ^2\s$ then
\s$l_ul_{u_1}=K\s$ and \s${dim}\s H^0(l_ul_{u_1})=2\s$ 
so that there always is a non-zero solution but
also \s$\theta(u)=0\s$ in this case due to the vanishing of the 
$2^{\rm nd}$ theta function on the r.h.s. of Eq.\s\s(\ref{Kk}).
Finally, if \s$u\pm u_1\s\not\in\s\NZ^2+\tau\NZ^2\s$ then
\s$s_2\in H^0(l_ul_{u_1})\s$ has to be proportional
to the element defined by (\ref{s_2}) and the condition
(\ref{soco}) coincides with the equation \s$\theta(u)=0\m$.
\vskip 0.9cm

\nappendix{2}
\vskip 0.5cm

Let us show that the 1,0-form $\mu$ satisfying relations 
(\ref{cfm}) and (\ref{cfm1}) automatically fulfills the condition
\qq
\int_{_\Sigma}{\kappa}\s\mu\wedge b\s=\s0\s.
\label{orth}
\qqq
Among the infinitesimal gauge field variations $\delta B$
given by Eq.\s\s(\ref{DB}) there are ones which are equivalent 
to infinitesimal gauge transformations:
\qq
\delta B\s=\s\de\Lambda\s+\s[B,\Lambda]\s.
\nonumber
\qqq
Explicitly, for \s$\Lambda=(\matrix{_{-\sigma}&_\varphi\cr^\kappa
&^\sigma})\s$ with $\sigma$ a function, $\varphi$ a section 
of $l_{u_1}^{-2}$ and $\kappa$ a section of $l_{u_1}^2$, 
this requires that
\qq
\de\kappa\s=\s0\s,\quad\quad
\pi\s\delta u_1\m({\rm Im}\m\tau)^{-1}\bar\omega\s=\s
-\de\sigma+\kappa\m b\s,\quad\quad\delta b\s=\s\de\varphi
+2\m\sigma\m b\s.
\label{three}
\qqq
Such variations may only change the normalization
of the theta function $\theta$. Integrating the second of 
the above relations against forms $\omega^a$ and using 
Eq.\s\s(\ref{kob}) we find that 
\qq
\delta u_1^a\s=\s-{_1\over^{2\pi i}}\s\epsilon^{ab}
\da_{b}\theta(u_1)
\label{du0}
\qqq
for the proper normalization of ${\kappa}$. For such $\delta u_1$
the first term on the right hand side 
of Eq.\s\s(\ref{nd}) gives a theta
function vanishing at $u=u_1$ and may be compensated by the
second term. The 3$^{\rm rd}$ equation of (\ref{three})
gives the compensating $\delta b\in\wedge^{01}(l_{u_1}^{-2})$.
Pairing Eq.\s\s(\ref{nd}) with the above $\delta u_1$ 
and $\delta b$ with the linear form $\phi$, we obtain 
the identity
\qq
{_1\over^i}\s\epsilon^{ab}
\da_{b}\theta(u_1)\s({\rm Im}\m\tau)_{ac}^{-1}
\int_{_\Sigma}\chi^c\wedge b\s+\s2\int_{_\Sigma}
\sigma\s\eta\wedge b\ =\ 0\ .
\label{cri}
\qqq
On the other hand,
\qq
\int_{_\Sigma}{\kappa}\s\mu\wedge b\s=\s
\int_{_\Sigma}\mu\wedge\de\sigma\s-\s{_1\over^{2i}}
\s\epsilon^{ab}\da_b\theta(u_1)\s({\rm Im}\m)^{-1}_{ac}
\int_{_\Sigma}\mu\wedge\bar\omega^c\s\cr
=\s-\int_{_\Sigma}\sigma\s\eta\wedge b\s-\s
{_1\over^{2i}}\s\epsilon^{ab}\da_b\theta(u_1)\s({\rm Im}
\m)^{-1}_{ac}\int_{_\Sigma}\chi^c\wedge b\ =\ 0
\nonumber
\qqq
where we have subsequently used the 2$^{\rm nd}$ equation
in (\ref{three}) with $\delta u_1$ given by Eq.\s\s(\ref{du0}), 
the relation $\de\mu=-\eta\wedge b$ and Eq.\s\s(\ref{cfm})
fixing $\mu$ and, finally, the identity (\ref{cri}).
\vskip 0.9cm

\nappendix{3}
\vskip 0.5cm

It is not difficult to see that there exist a non-zero 
element $P\in S^4 H^0(L_\Theta^2)$, a homogeneous polynomial 
of degree 4 on $H^0(L_\Theta^2)^*$, s.t. 
\qq
P(\phi_{u'})\s=\s0
\nonumber
\qqq
for all $u'\in\NC^2$. Indeed, $dim\s S^4H^0(L_\Theta^2)
=({7\atop3})=35$ but the map $u'\mapsto P(\phi_{u'})$ 
defines an even theta function of order 8 and 
$dim\s H^0_{\rm even}(L_\Theta^8)=34$. 
$P$ is a quartic expression in $\theta_{2,e}(u')$
which vanishes for all $u'$. It has to be preserved
by the $(\NZ/2\NZ)^4$-action (\ref{acone}) and
hence it must be of the form
\qq
P&=&c_1\s(\theta_{2,(0,0)}^{\s4}+\s\theta_{2,(1,0)}^{\s4}
+\s\theta_{2,(0,1)}^{\s4}+\s\theta_{2,(1,1)}^{\s4})\cr\cr
&+&c_2\s(\theta_{2,(0,0)}^{\s2}\s\m\theta_{2,(1,0)}^{\s2}+
\s\theta_{2,(0,1)}^{\s2}\s\m\theta_{2,(1,1)}^{\s2})\cr\cr
&+&c_3\s(\theta_{2,(0,0)}^{\s2}\s\m\theta_{2,(0,1)}^{\s2}+
\s\theta_{2,(1,0)}^{\s2}\s\m\theta_{2,(1,1)}^{\s2})\cr\cr
&+&c_4\s(\theta_{2,(0,0)}^{\s2}\s\m\theta_{2,(1,1)}^{\s2}+
\s\theta_{2,(1,0)}^{\s2}\s\m\theta_{2,(0,1)}^{\s2})\cr\cr
&+&c_5\s\theta_{2,(0,0)}\s\m\theta_{2,(1,0)}\s\m\theta_{2,(0,1)}
\s\m\theta_{2,(1,1)}\s.
\nonumber
\qqq
It is not difficult to calculate the values of coefficients
$c_i$. Denoting $\alpha\equiv\theta_{2,(0,0)}(0)$,
\s$\beta\equiv\theta_{2,(1,0)}(0)$, 
\s$\gamma\equiv\theta_{2,(0,1)}(0)\s$
and \s$\delta\equiv\theta_{2,(1,1)}(0)$,
one has
\qq
c_1&=&\ \ \m(\alpha^2\beta^2-\gamma^2\delta^2)
(\alpha^2\gamma^2-\beta^2\delta^2)(\alpha^2\delta^2
-\beta^2\gamma^2)\s,\cr\cr
c_2&=&-(\alpha^4+\beta^4-\gamma^4-\delta^4)
(\alpha^2\gamma^2-\beta^2\delta^2)(\alpha^2\delta^2
-\beta^2\gamma^2)\s,\cr\clabel{em}\cr
c_3&=&-(\alpha^4-\beta^4+\gamma^4-\delta^4)
(\alpha^2\beta^2-\gamma^2\delta^2)(\alpha^2\delta^2
-\beta^2\gamma^2)\s,\cr\cr
c_4&=&-(\alpha^4-\beta^4-\gamma^4+\delta^4)
(\alpha^2\beta^2-\gamma^2\delta^2)(\alpha^2\gamma^2
-\beta^2\delta^2)\s,\cr\cr
c_5&=&\ \ \m2\m \alpha\beta\gamma\delta\m
[(\alpha^4-\beta^4+\gamma^4-\delta^4)^2-\m 
4(\alpha^2\gamma^2-\beta^2\delta^2)^2]\s.
\qqq
If we use the basis dual to $(\theta_{2,e})$ to
identify $\phi\in H^0(L_\Theta^2)^*$ with a vector
$p=(p_1,p_2,p_3,p_4)\in\NC^4$, the equation
of the Kummer quartic $\CK^*$ becomes
\qq
c_1\s(p_1^4+p_2^4+p_3^4+p_4^4)\s+\s c_2\s
(p_1^2\m p_2^2\s+\s p_3^2\m p_4^2)
\s+\s c_3\s(p_1^2\m p_3^2\s+\s p_2^2\m p_4^2)\s\m\cr
\clabel{kuma}\cr
+\s c_4\s(p_1^2\m p_4^2\s+\s p_2^2\m p_3^2)
\s+\s c_5\s p_1\m p_2\m p_3\m p_4\ =\ 0\s.
\qqq
Similarly, identifying $\theta\in H^0(L_\Theta^2)$ with
$q=(q_1,q_2,q_3,q_4)\in\NC^4$ with the help of the basis
$(\theta_{2,e})$, the same equation with $p$ replaced by
$q$ defines the Kummer quartic $\CK$, compare \cite{Kummer}, 
page 81.
\vskip 0.4cm

We shall also need another well known presentation of the above
equation using the inhomogeneous coordinates of the Weierstrass 
points $\lambda_s$ given by Eq.\s\s(\ref{wp}). It is usually
obtained by beautiful geometric considerations about quadratic
line complexes, see \cite{GH}. It may be also obtained
analytically by observing that the multivalued functions
\qq
x\ \mapsto\ \theta_{2,e}(\smallint_{x_0}^x\omega-\Delta)
\nonumber
\qqq
transform like bilinears in \s$\da_a\vartheta(
\smallint_{x_0}^x\omega-\Delta)\m$, \s i.e.\s\s that they
represent quadratic differentials. It follows that
\qq
\sum\limits_e\theta_{2,e}(E_s)\s\s\theta_{2,e}(
\smallint_{x_0}^x\omega-\Delta)\ =\ \vartheta(E_s+
\smallint_{x_0}^x\omega-\Delta)\s\s\vartheta(E_s-
\smallint_{x_0}^x\omega+\Delta)\s\cr
=\ D_s\s\s\left(\da_1\vartheta(E_s')\s\m\da_2\vartheta(
\smallint_{x_0}^x\omega-\Delta)\s-\s
\da_2\vartheta(E_s')\s\m\da_1\vartheta(
\smallint_{x_0}^x\omega-\Delta)\right)\s\clabel{rlti}\cr
\ \cdot\s\left(\da_1\vartheta(E_s'')\s\m\da_2\vartheta(
\smallint_{x_0}^x\omega-\Delta)\s-\s
\da_2\vartheta(E_s'')\s\m\da_1\vartheta(
\smallint_{x_0}^x\omega-\Delta)\right)\s
\qqq
where $E_s=\hf(e_s+\tau e'_s)\s$ is an odd characteristics
from the list (\ref{ordr}) and $E_s'\m,\s E_s''$ are the
two other ones s.t. $E_s+E_s'=E_s''\s\m{\rm mod}\m(\NZ^2+\tau\NZ^2)$.
The odd characteristics $E_s,\m E_s',\m E_s''$ are
either a permutation of $E_1,\m E_4,\m E_5$ or 
a permutation of $E_2,\m E_3,\m E_6$.
The relations (\ref{rlti}) hold since both sides represent
a quadratic differential with double zeros at the Weierstrass
points corresponding to $E_s'$ and $E_s''$. One may obtain
expressions for the coefficients $D_s$ by the de l'Hospital
rule applied twice at those points. Specifying then
$\smallint_{x_0}^x\omega-\Delta$ to $E_s$ or to 3 remaining
odd characteristics one obtains relations for
quadratic combinations of $\theta_{2,e}(0)$ of the form
\s$\pm\alpha^2\pm\beta^2\pm\gamma^2\pm\delta^2\s$
with 2 plus and 2 minus signs as well as for
\s$\alpha\beta\pm\gamma\delta\m$, 
\s$\alpha\gamma\pm\beta\delta\m$
and \s$\alpha\delta\pm\beta\gamma\m$. 
These relations may be used to compute the ratios 
of the coefficients $c_i$ (\ref{em})
which become functions of $\lambda_s$ only.
One obtains this way an alternative expression  
for the coefficients $c_i$ 
\qq
c_1&=&(\lambda_1-\lambda_2)(\lambda_3-\lambda_4)
(\lambda_5-\lambda_6)\s,\cr\cr
c_2&=&2(\lambda_1-\lambda_2)
((\lambda_3-\lambda_5)(\lambda_4-\lambda_6)
+(\lambda_3-\lambda_6)(\lambda_4-\lambda_5))\s,\cr\cr
c_3&=&-2(\lambda_3-\lambda_4)((\lambda_1-\lambda_5)
(\lambda_2-\lambda_6)
+(\lambda_1-\lambda_6)(\lambda_2-\lambda_5))\s,
\clabel{kumb}\cr\cr
c_4&=&2(\lambda_5-\lambda_6)((\lambda_1-\lambda_3)
(\lambda_2-\lambda_4)
+(\lambda_1-\lambda_4)(\lambda_2-\lambda_3))\s,\cr\cr
c_5&=&-2(\lambda_1-\lambda_3)((\lambda_4-\lambda_5)
(\lambda_2-\lambda_6)
+(\lambda_4-\lambda_6)(\lambda_2-\lambda_5))\cr
&&-2(\lambda_1-\lambda_4)((\lambda_3-\lambda_5)
(\lambda_2-\lambda_6)
+(\lambda_3-\lambda_6)(\lambda_2-\lambda_5))\cr
&&-2(\lambda_1-\lambda_5)((\lambda_2-\lambda_4)
(\lambda_3-\lambda_6)
+(\lambda_2-\lambda_3)(\lambda_4-\lambda_6))\cr
&&-2(\lambda_1-\lambda_6)((\lambda_2-\lambda_4)
(\lambda_3-\lambda_5)
+(\lambda_2-\lambda_3)(\lambda_4-\lambda_5))\s.
\qqq
equivalent to the previous one up to normalization.
Note that the $SL(2,\NC)$ transformations 
\s$\lambda_s\mapsto{a\lambda_s+b\over c\lambda_s+d}\s$ 
preserve the form the quartic equation.
The virtue of the analytic approach is that it
also provides useful expressions for the non-homogeneous 
ratios like e.g.
\qq
{\alpha\beta+\gamma\delta\over\alpha^2\gamma^2-\beta^2\delta^2}
\ =\ -\s{\ee^{-\m{1\over 2}\m \pi i\s(1,0)\cdot\tau\m(1,0)}\over
2\s\m C^2\s\s(\da_2\vartheta(E_1))^2}\ {(\lambda_2-\lambda_5)
(\lambda_2-\lambda_6)(\lambda_3-\lambda_4)\over\lambda_1-\lambda_2}\s.
\label{tbus}
\qqq
$C^2$ is given by the equations
\qq
C^2\s=\s\hf\s{_{(\da_1\vartheta)^3\s\da_2^3\vartheta
\s-\s 3\s(\da_1\vartheta)^2\s\da_2\vartheta
\s\da_1\da_2^2\vartheta
\s+\s 3\s\da_1\vartheta\s(\da_2\vartheta)^2
\s\da_1^2\da_2\vartheta
\s-\s(\da_2\vartheta)^3
\s\da_1^3\vartheta}\over^{(\da_2\vartheta)^4}}\bigg\vert_{E_s}
\ \prod\limits_{t\s\not=\s s}(\lambda_s-\lambda_t)\hspace{0.4cm}
\nonumber
\qqq
holding for any fixed $s$. It is not difficult to see by 
differentiating twice Eq.\s\s(\ref{inho}) at $x=x_s$ 
that $C$ is the same constant that appears in Eq.\s\s(\ref{newf}). 
The expression (\ref{tbus}) is used below to fix the 
normalization of the Hitchin map.
\vskip 0.9cm

\nappendix{4}
\vskip 0.5cm

We shall show here that the overall normalization of the Hitchin map
is as in Eq.\s\s(\ref{glr}). Since 
\qq
&&\ee^{\m\pi i\s e_1'\cdot\tau\m e_1'\s+\s 4\pi i\s e_1'
\cdot u_1}\s\s\vartheta(2u_1+E_1)^2\s\cr\cr
&&=\s-\m\ee^{\m\pi i\s e_1'\cdot\tau\m e_1'}
\s\s\vartheta(2u_1+E_1)\s\s\vartheta(2u_1-E_1)
=\s-\m\ee^{\m\pi i\s e_1'\cdot\m\tau\m e_1'}\s\sum\limits_e
\theta_{2,e}(E_1)\s\s\theta_{2,e}(2u_1)\s\cr
&&=\s-\m\ee^{\m{1\over 2}\m\pi i\s(1,0)\cdot\tau(1,0)}
\s\sum\limits_e(-1)^{(1,0)\cdot e}\s\s\theta_{2,e+(1,0)}(0)
\s\s\theta_{2,e}(2u_1)\s,
\nonumber
\qqq
the coefficient of ${(d\lambda)^2\over\zeta^2}$ 
on the right hand side of Eq.\s\s(\ref{123}) takes 
at $\lambda=\lambda_2$ the value
\qq
\m{_{C^2}\over^{16\pi^2}}\s\s\ee^{\m{1\over 2}\m\pi i\s(1,0)
\cdot\tau(1,0)}\ (\da_2\vartheta(E_1))^2\ 
(\lambda_1-\lambda_2)^2\s\s(\beta\m\theta_{2,(0,0)}(2u_1)
\s-\s\alpha\m\theta_{2,(1,0)}(2u_1)\cr
+\s\delta\m\theta_{2,(0,1)}(2u_1)\s
-\s\gamma\m\theta_{2,(1,1)}(2u_1))
\label{ha}
\qqq
in the notations of Appendix 3.
This coefficient should coincide with the one obtained from the 
right hand side of Eq.\s\s(\ref{glr}) which is equal to
\qq
-\m{_1\over^{64\pi^2}}\sum\limits_{t\s\not=\s2}r_{2\m t}(q,p)\s
\prod\limits_{v\s\not=\s2,t}(\lambda_2-\lambda_v)
\label{ho}
\qqq
calculated at $(q,p)$ corresponding to $(\theta,\phi_{u_1})$
with $\theta$ given by Eq.\s\s(\ref{121}). The respective
values of $r_{st}$ are:
\qq
r_{1\m t}&=& 0\s,\cr\cr
r_{23}&=&2\s(-\m\alpha\gamma^2
\m\theta_{2,(0,0)}(2u_1)\s-\s\beta\delta^2\s\theta_{2,(1,0)}(2u_1)
\s-\s\gamma\alpha^2\s\theta_{2,(0,1)}(2u_1)\cr
&&-\s\delta\beta^2\s\theta_{2,(1,1)}(2u_1)
\s-\s\beta\gamma\delta\s\theta_{2,(0,0)}(2u_1)
\s-\s\alpha\gamma\delta\s\theta_{2,(1,0)}(2u_1)\cr
&&-\s\alpha\beta\delta\s\theta_{2,(0,1)}(2u_1)
\s-\s\alpha\beta\gamma\s\theta_{2,(1,1)}(2u_1)\m)\s,\cr\cr
r_{24}&=&2\s(\m\alpha\gamma^2
\s\theta_{2,(0,0)}(2u_1)\s+\s\beta\delta^2\s\theta_{2,(1,0)}(2u_1)
\s+\s\gamma\alpha^2\s\theta_{2,(0,1)}(2u_1)\cr
&&+\s\delta\beta^2\s\theta_{2,(1,1)}(2u_1)
\s-\s\beta\gamma\delta\s\theta_{2,(0,0)}(2u_1)
\s-\s\alpha\gamma\delta\s\theta_{2,(1,0)}(2u_1)\cr
&&-\s\alpha\beta\delta\s\theta_{2,(0,1)}(2u_1)
\s-\s\alpha\beta\gamma\s\theta_{2,(1,1)}(2u_1)\m)\s,\clabel{00}\cr\cr
r_{25}&=&2\s(\m\alpha\delta^2
\s\theta_{2,(0,0)}(2u_1)\s+\s\beta\gamma^2\s\theta_{2,(1,0)}(2u_1)
\s+\s\gamma\beta^2\s\theta_{2,(0,1)}(2u_1)\cr
&&+\s\delta\alpha^2\s\theta_{2,(1,1)}(2u_1)
\s+\s\beta\gamma\delta\s\theta_{2,(0,0)}(2u_1)
\s+\s\alpha\gamma\delta\s\theta_{2,(1,0)}(2u_1)\cr
&&+\s\alpha\beta\delta\s\theta_{2,(0,1)}(2u_1)
\s+\s\alpha\beta\gamma\s\theta_{2,(1,1)}(2u_1)\m)\s,\cr\cr
r_{26}&=&2\s(-\m\alpha\delta^2
\s\theta_{2,(0,0)}(2u_1)\s-\s\beta\gamma^2\s\theta_{2,(1,0)}(2u_1)
\s-\s\gamma\beta^2\s\theta_{2,(0,1)}(2u_1)\cr
&&-\s\delta\alpha^2\s\theta_{2,(1,1)}(2u_1)
\s+\s\beta\gamma\delta\s\theta_{2,(0,0)}(2u_1)
\s+\s\alpha\gamma\delta\s\theta_{2,(1,0)}(2u_1)\cr
&&+\s\alpha\beta\delta\s\theta_{2,(0,1)}(2u_1)
\s+\s\alpha\beta\gamma\s\theta_{2,(1,1)}(2u_1)\m)\s.
\nonumber
\qqq
Multiplying the coefficients at subsequent
$\theta_{2,e}(2u_1)$ in expression (\ref{ha}) 
by $\alpha,\s-\beta,\s\gamma$ and $-\delta$, respectively, 
and summing them up we obtain
\qq
\m{_{C^2}\over^{8\pi^2}}\s\s\ee^{\m{1\over 2}\m\pi i\s 
(1,0)\cdot\tau(1,0)} 
\ (\da_2\vartheta(E_1))^2\ 
(\lambda_1-\lambda_2)^2\s\s(\alpha\beta+\gamma\delta)\s.
\nonumber
\qqq
A similar operation on expression (\ref{ho}) gives
\qq
-\m{_1\over^{16\pi^2}}\s\s(\lambda_1-\lambda_2)
(\lambda_2-\lambda_5)
(\lambda_2-\lambda_6)(\lambda_3-\lambda_4)
\s(\alpha^2\gamma^2-\beta^2\delta^2)\s.
\nonumber
\qqq
The equality of the two expressions follows from 
Eq.\s\s(\ref{tbus}). This verifies the correctness
of the overall normalization of the Hitchin map in 
Eq.\s\s(\ref{glr}).

\end{document}